\documentclass[prd,aps,nofootinbib,preprint,tightenlines]{revtex4}
\newcommand{\checked}[1]{}

\usepackage{float}
\usepackage{extarrows}
\usepackage{epsfig}
\usepackage{color}
\usepackage{tabularx}
\usepackage{slashbox}
\usepackage{diagbox}
\usepackage{graphicx}
\usepackage{booktabs}
\usepackage{hyperref}
\hypersetup{
    colorlinks=true,
    linkcolor=blue,
    citecolor=blue,
    filecolor=blue,
    urlcolor=blue,
}

\newcommand{\be}{\begin{equation}}
\newcommand{\ee}{\end{equation}}
\newcommand{\ba}{\begin{eqnarray}}
\newcommand{\ea}{\end{eqnarray}}

\newcommand{\la}{\label}

\newcommand{\cQ}{{\cal Q}}

\newcommand{\e}{\epsilon}

\begin{document}

\title{Real-time hard-thermal-loop gluon self-energy in a semiquark-gluon plasma}

\author{Yubiao Wang$^{a}$, Qianqian Du$^{a,b}$, and
Yun Guo$^{a,b}\footnote{yunguo@mailbox.gxnu.edu.cn}$}
\affiliation{$^a$Department of Physics, Guangxi Normal University, Guilin, 541004, China\\
$^b$Guangxi Key Laboratory of Nuclear Physics and Technology, Guilin, 541004, China
\vspace*{2cm}
}

\begin{abstract}
In the real-time formalism of the finite-temperature field theory, we compute the one-loop gluon self-energy in a semiquark-gluon plasma (QGP) where a background field $\cQ$ has been introduced for the vector potential, leading to a nontrivial expectation value for the Polyakov loop in the deconfined phase. 
Explicit results of the gluon self-energies up to the next-to-leading order in the hard-thermal-loop approximation are obtained. We find that for the retarded/advanced gluon self-energy, the corresponding contributions at next-to-leading order are formally analogous to the well-known result at $\cQ=0$ where the background field modification on the Debye mass is entirely encoded in the second Bernoulli polynomials. The same feature is shared by the leading order contributions in the symmetric gluon self-energy where the background field modification becomes more complicated, including both trigonometric functions and the Bernoulli polynomials. These contributions are nonvanishing and reproduce the correct limit as $\cQ \rightarrow 0$. In addition, the leading order contributions to the retarded/advanced gluon self-energy and the next-to-leading order contributions to the symmetric gluon self-energy are completely new as they only survive at $\cQ\neq0$. Given the above results, we explicitly verify that the Kubo-Martin-Schwinger condition can be satisfied in a semi-QGP with a nonzero background field.
\end{abstract}
\maketitle
\newpage

\section{Introduction}\la{intro}

The nature of the quark-gluon plasma (QGP), a primordial state of matter generated in ultrarelativistic heavy-ion experiments has been  systematically studied over the last decades. Understanding the deconfining phase transition from the normal hadronic matter to the QGP is one of the most important goals in high-energy nuclear physics. Near the critical temperature $T_c$, a tough challenge to achieve such a goal emerges because of the failure of the perturbation theory based on weak coupling expansion\cite{Kapusta:1979fh,Kajantie:2002wa,Andersen:2009tc}. In a region from $T_c$ to $\sim 4 T_c$, numerical simulations on the lattice provide a powerful tool to study the nonperturbative physics in the partially deconfined system which is termed a semi-QGP\cite{Hidaka:2008dr}. Although the wealth of information on the thermodynamics in equilibrium has been obtained from lattice QCD\cite{Boyd:1996bx,Borsanyi:2012ve,Panero:2009tv,Bazavov:2009zn,Borsanyi:2010cj}, due to the well-known sign problem, exploring the equation of state at very large baryon chemical potential remains to be solved.

An alternative solution is to develop effective theories to investigate the properties of the strongly interacting matter. As shown in lattice simulations for pure $SU(N)$ gauge theories, a significant increase of the order parameter for deconfinement was observed in the semi-QGP region where the Polyakov loop is nonzero but less than unity\cite{Kaczmarek:2002mc,Petreczky:2004pz,Gupta:2007ax}. Such a notable feature can be described by considering a classical background field $A_0^{\rm cl}$ for the vector potential which is a diagonal matrix in the color space, $(A^{\rm cl}_0)_{ab}= \delta_{ab} \cQ^a/g $ with the matrix elements satisfying $\sum_{a=1}^{N} \cQ^a=0$ for $SU(N)$ gauge group. Accordingly, the effective potential or free energy can be computed perturbatively by using a constrained path integral\cite{Belyaev:1991gh,Bhattacharya:1992qb,KorthalsAltes:1993ca,KorthalsAltes:1999cp,Dumitru:2013xna,Reinosa:2015gxn,Maelger:2017amh,Guo:2018scp}. The resulting effective potential attains at the minimum at vanishing background field, indicating a deconfining phase at all temperatures. To drive the system to confinement, nonperturbative contributions have been introduced in the matrix models that generate a complete repulsion of eigenvalues of the thermal Wilson line. In recent years, much attention has been paid to the developments of the matrix models which have already had great success in studying the QCD phase transition\cite{Meisinger:2001cq,Dumitru:2010mj,Dumitru:2012fw,Pisarski:2012bj,Guo:2014zra,Pisarski:2016ixt}.

In the meanwhile, many phenomenological applications have been considered in the semi-QGP where the focus was put on the influence of the background field on the corresponding physical quantities such as transport coefficients and electromagnetic probes, see Refs.~\cite{Hidaka:2009ma,Gale:2014dfa,Hidaka:2015ima,Singh:2018wps,Singh:2019cwi,Lin:2013efa} for examples. In this work, we concentrate on the real-time gluon self-energy in the presence of a background field. Being crucial in many processes involving soft momentum exchange, the gluon self-energy has been extensively studied in previous works where a widely used calculational technique is the so-called Hard-Thermal-Loop (HTL) approximation\cite{Braaten:1989mz}. 
Besides the HTL gluon self-energies in equilibrium\cite{Bellac:2011kqa}, the investigation on the viscous corrections shows the existence of unstable modes of the plasma due to a rapid exponential growth of the soft gluon fields\cite{Romatschke:2003ms, Romatschke:2004jh,Du:2016wdx}. In addition, the HTL gluon self-energy in a semi-QGP has been computed in the imaginary time formalism\cite{Hidaka:2009hs}. 
The most surprising outcome is that there is an anomalous contribution $\sim T^3$ appearing in the gluon self-energy that vanishes at zero background field.

In the real-time formalism, the gluon self-energy becomes a $2\times 2$ matrix. The four components are not independent, therefore, it is more useful to write the gluon self-energy in terms of the three independent components in Keldysh representation\cite{Keldysh:1964ud}, namely,
\be
\Pi_R=\Pi_{11}+\Pi_{12}\, ,\quad \Pi_A=\Pi_{11}+\Pi_{21}\, ,\quad \Pi_F=\Pi_{11}+\Pi_{22}\, .
\ee
In the absence of the background field, by analytically continuing the result in imaginary time, one can easily get the retarded and advanced gluon self-energies as denoted by $\Pi_R$  and $\Pi_A$, respectively. Furthermore, the symmetric gluon self-energy $\Pi_F$ in an equilibrium QGP can be obtained via the Kubo-Martin-Schwinger (KMS) condition\cite{Kubo:1957mj, Martin:1959jp}
\be\la{kmseq}
\Pi_F(P) = \big(1+2 n(p_0)\big) {\rm sgn} (p_0) \big(\Pi_{R}(P)-\Pi_{A}(P) \big)\, ,
\ee
where $n(p_0)$ is the Bose-Einstein distribution function. When a nonzero background field is considered, a $\cQ$-dependent modification on the distribution function needs to be taken into account. Furthermore, besides the normal $\sim T^2$ terms, the aforementioned anomalous contribution $\sim T^3$ remains in the retarded/advanced gluon self-energy after analytical continuation. As a result, one cannot expect a trivial extension of Eq.~(\ref{kmseq}) from $\cQ=0$ to $\cQ \neq 0$. Therefore, the computation of the real-time gluon self-energies as well as the verification of the KMS condition in a semi-QGP with a nonzero background field will be the main concern in the current work.

The rest of the paper is organized as follows. In Sec.~\ref{feynman}, we briefly review the double line basis as commonly used when computing in a background field and summarize the corresponding Feynman rules in the real-time formalism. In Sec.~\ref{pire}, we compute the one-loop retarded/advanced gluon self-energy in a semi-QGP up to the next-to-leading order (NLO) in the HTL approximation and discuss the $\cQ$ modifications on the Debye screening mass. The same analysis on the symmetric gluon self-energy is carried out in Sec.~\ref{pisy}. With the obtained results, in Sec.~\ref{kms} we explicitly verify that the KMS condition can be satisfied order by order by the HTL approximated gluon self-energies in a semi-QGP where $\cQ \neq 0$. Conclusions and outlook are given in Sec.~\ref{conclusions}. In addition, some details of the calculation of the contributions from the pure gauge part can be found in Appendix~\ref{app}. A reorganization of the LO contributions of the symmetric gluon self-energy is discussed in Appendix~\ref{repif}.

\section{The Feynman rules in the double line basis}\la{feynman}

For completeness, we briefly review the double line basis\cite{Hidaka:2009hs,Cvitanovic:1976am} which is defined by
the generators of the fundamental representation,
\be
\la{21}
(t^{ab})_{cd} = \frac{1}{\sqrt{2}}{\cal P}^{ab}_{cd}\, ,
\ee
where the projection operator is given by
\be\la{22}
{\cal P}^{ab}_{cd} ={\cal P}^{ab,dc}={\cal P}_{ba,cd}  = \delta^{a}_{c}\delta^{b}_{d}-\frac{1}{N}\delta^{ab}\delta_{cd}\, .
\ee
The normalization of generators reads
\be
\textrm{tr}(t^{ab}\,t^{cd})= \frac{1}{2}{\cal P}^{ab, cd}\, .
\ee
For $SU(N)$ gauge theories, the color indices $a,b,c$ and $d$ run from $1$ to $N$. There are $N^2-N$ off-diagonal generators $t^{ab}$ with $a\neq b$ which are the ladder operators of the Cartan basis. They are orthogonal to each other and normalized as $\textrm{tr}(t^{ab}\, t^{ba})=1/2$ with fixed $a$ and $b$. The $N$ diagonal generators are not independent, satisfying $\sum_{a=1}^N t^{aa}=0$ and the normalization becomes $\textrm{tr}(t^{aa}\,t^{bb})= (\delta^{ab}-1/N)/2$ where no summation over $a$ and $b$ applies.

The great advantage of computation in the over-complete double line basis is that the classical covariant derivative $D_{\mu}^{\rm {cl}}$ acting upon the fields has a very simple form in momentum space. In the fundamental representation, $D_{\mu}^{\rm {cl}}=\partial _\mu-i g A_{\mu}^{\rm {cl}}$ with $A_{\mu}^{\rm {cl}} =A_{0}^{\rm {cl}} \delta_{\mu 0}$. When this covariant derivative acts upon a quark field, we have $D_0^{{\rm cl}}\, \psi_a \rightarrow -i (p_0+\cQ^a) \psi_a$. 
Similarly, in the adjoint representation, $D_{\mu}^{\rm {cl}}=\partial _\mu-i g [A_{\mu}^{\rm {cl}}, \, \cdots]$. Acting upon a bosonic field, it gives $D_0^{{\rm cl}}\, t^{ab} \rightarrow -i (p_0+\cQ^a-\cQ^b) t^{ab}$.  
In any case, there is only a constant and color-dependent shift in the energies.

Given the QCD Lagrangian $ {\cal L}_{\rm QCD}$, we expand the gauge fields around some fixed classical values as $A_\mu=A^{\rm cl}_\mu + B_\mu$ where $B_\mu$ corresponds to the quantum fluctuation. With the standard procedure, the corresponding Feynman rules in the double line basis can be derived. For example, the inverse bare gluon propagator in momentum space reads
\be\la{inbare1}
\frac{\delta {\cal S}}{\delta B_\mu^{ba}(P) \delta B_\nu^{dc}(-P)}=\Big( (P^{ab})^2\delta_{\mu\nu}-(1-\frac{1}{\xi})P_\mu^{ab}P_\nu^{ab}\Big){\cal P}^{ab,cd}\, ,
\ee
where the action ${\cal S} = \int d^4 x \,  {\cal L}_{\rm QCD}$ and $\xi$ is the gauge fixing parameter. The $\cQ$-dependent momentum $P^{ab}$ is given by $P^{ab}=(p_0+\cQ^a-\cQ^b, {\bf{p}})$. Notice that due to the over-completeness of the basis, upon inversion,  the explicit form of the diagonal components of the gluon propagator cannot be uniquely determined. However, such an ambiguity is absent in the calculation of the gluon self-energy after performing the sum over the color indices\cite{Guo:2020jvc}.

The complete Feynman rules in the imaginary time formalism can be found in Ref.~\cite{Hidaka:2009hs}. On the other hand, to compute in the real-time formalism, one has to double the field degrees of freedom so that the propagators become a $2 \times 2$ matrix. In order to get the Feynman rules in the presence of the background field, we follow Refs.~\cite{Hidaka:2009ma, Furuuchi:2005zp} and take the background field only for the part in imaginary time (not in real time) along a complex time path\cite{Bellac:2011kqa}. As a result, the corresponding Feynman rules become very simple because the background field acting as an imaginary chemical potential only affects the statistical distributions of the thermal partons.

\begin{figure}
\centering
\includegraphics[width=0.85\textwidth]{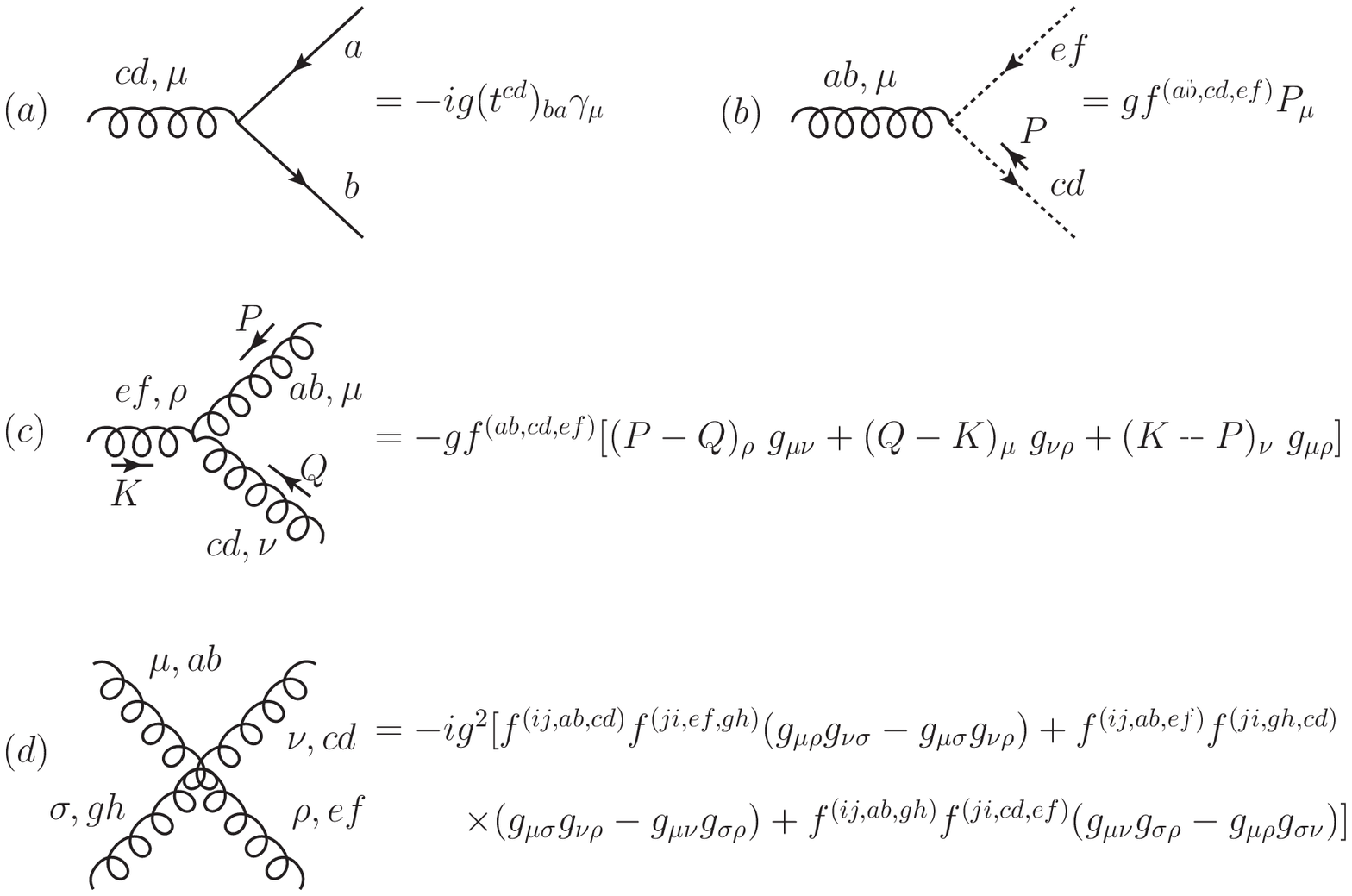}
\vspace*{-0.2cm}
\caption{\label{fey}
Feynman rules for vertices in the double line notation where the structure constant is given by $f^{ab,cd,ef}=\frac{i}{\sqrt{2}}(\delta^{ad}\delta^{ad}\delta^{ad}-\delta^{ad}\delta^{ad}\delta^{ad})$.}
\end{figure}

Let us denote the bare gluon propagator as $G^{ab,cd}_{\mu \nu}(K,\cQ^{ab})=-g_{\mu \nu}{\cal P}^{ab,cd}D(K,\cQ^{ab})$ with no summation over $a$ or $b$ and $\cQ^{ab}\equiv \cQ^a-\cQ^b$. Here, $D(K,\cQ^{ab})$ is a $2\times 2$ matrix which reads
\ba
D(K,\cQ^{ab})=&& \left (\begin{array}{cc}
\frac{i}{K^2+i\epsilon} & 0\\
0 & \frac{-i}{K^2-i\epsilon} \\
       \end{array} \right )
       + 2 \pi \delta(K^2)
\left (\begin{array}{cc}
n(k_0,\cQ^{ab}) &  n(k_0,\cQ^{ab})+\theta(-k_0) \\
n(k_0,\cQ^{ab})+\theta(k_0) & n(k_0,\cQ^{ab})  \\
\end{array} \right ) \, , \nonumber \\
\ea
where $\theta(k_0)$ is the Heaviside step function and the $\cQ$-dependent distribution function is defined as
\begin{equation}\la{bd}
n(k_0,\cQ^{ab})=\left\{
\begin{aligned}
&\frac{1}{e^{(|k_0|-i \cQ^{ab})/T}-1}\equiv n_{+}^{ab}(k_0) \quad\quad {\rm for}\quad k_0>0\\
&\frac{1}{e^{(|k_0|+i\cQ^{ab})/T}-1}\equiv n_{-}^{ab}(k_0) \quad\quad {\rm for}\quad k_0<0 \\
\end{aligned}
\right.\,.
\end{equation}
In the Keldysh representation, the retarded, advanced and symmetric propagator can be obtained as the following
\ba
D_{R/A}(K)&=&D_{11}(K,\cQ^{ab})-D_{12}(K,\cQ^{ab})=\frac{i}{K^2\pm i\, {\rm sgn}(k_0)\epsilon}\, ,\nonumber \\
D_{F}(K,\cQ^{ab})&=&D_{11}(K,\cQ^{ab})+D_{22}(K,\cQ^{ab})=2\pi(1+2n(k_0,\cQ^{ab}))\delta(K^2)\, .
\ea
Here, $+$ and $-$ correspond to the retarded and advanced propagators, respectively and ${\rm sgn}(k_0)$ is the sign function. As we can see, the nonzero background field alters only the symmetric propagator.

The bare quark propagator can be obtained in a similar way which we denote as ${\tilde G}^{ab}(K,\cQ^a)=\displaystyle{\not} K\delta^{ab}{\tilde D}(K,\cQ^{a})$ with no summation over $a$. The matrix elements of ${\tilde D}(K,\cQ^{a})$ read
\ba
{\tilde D}(K,\cQ^{a})=&& \left (\begin{array}{cc}
\frac{i}{K^2+i\epsilon} & 0\\
0 & \frac{-i}{K^2-i\epsilon} \\
       \end{array} \right )
       + 2 \pi \delta(K^2)
\left (\begin{array}{cc}
{\tilde n}(k_0,\cQ^{a}) &  {\tilde n}(k_0,\cQ^{a})+\theta(-k_0) \\
{\tilde n}(k_0,\cQ^{a})+\theta(k_0) & {\tilde n}(k_0,\cQ^{a})  \\
\end{array} \right ) \, , \nonumber \\
\ea
where the fermionic distribution function is given by
\begin{equation}\la{fd}
{\tilde n}(k_0,\cQ^{a})=\left\{
\begin{aligned}
&\frac{1}{e^{(|k_0|-i\cQ^{a})/T}+1}\equiv {\tilde n}_{+}^{a}(k_0) \quad\quad {\rm for}\quad k_0>0\\
&\frac{1}{e^{(|k_0|+i\cQ^{a})/T}+1}\equiv {\tilde n}_{-}^{a}(k_0) \quad\quad {\rm for}\quad k_0<0 \\
\end{aligned}
\right.\,.
\end{equation}
Accordingly, the three independent components in the Keldysh representation take the following forms:
\ba
{\tilde D}_{R/A}(K)&=&{\tilde D}_{11}(K,\cQ^{a})-{\tilde D}_{12}(K,\cQ^{a})=\frac{i}{K^2\pm i\, {\rm sgn}(k_0)\epsilon}\, ,\nonumber \\
{\tilde D}_{F}(K,\cQ^a)&=&{\tilde D}_{11}(K,\cQ^{a})-{\tilde D}_{12}(K,\cQ^{a}) = 2\pi(1-2{\tilde n}(k_0,\cQ^{a}))\delta(K^2)\, .
\ea

In addition, the ghost propagator can be obtained from the gluon propagator by dropping the metric tensor $-g_{\mu\nu}$. The Feynman rules for various vertices are listed in Fig.~\ref{fey}. For later use, we also define the periodic Bernoulli polynomials,
\ba
B_{2l}(x)=\sum_{n=1}^\infty (-1)^{l-1} \frac{2 (2l)!}{(2\pi n)^{2l}}\textrm{cos}(2\pi x n)\, ,
\ea
which satisfy
\ba
2l B_{2l-1}(x)=\frac{d }{d x} B_{2l}(x)\, .
\ea
It is easy to show that the above defined Bernoulli polynomials are periodic functions of $x$, with period $1$. For $0\le x \le 1$, the second and third Bernoulli polynomials which are relevant in this work read
\be\la{aa3}
B_2(x)=x^2-x+\frac{1}{6}\, ,\quad B_3(x)=x^3-{\frac{3}{2}}x^2+\frac{1}{2}x\, .
\ee
For arbitrary values of $x$, the argument of the above Bernoulli polynomials should be understood as $x-[x]$ with $[x]$ the largest integer less
than $x$, which is nothing but the modulo function.

\begin{figure}[hbtp]
\centering
\includegraphics[width=0.8\textwidth]{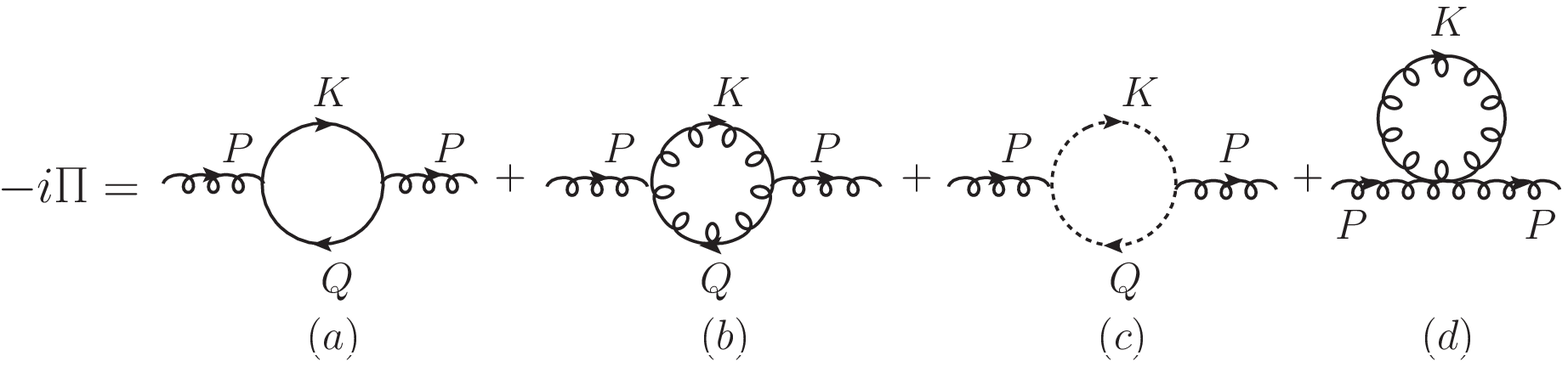}
\vspace*{-0.2cm}
\caption{\label{gse}
Feynman diagrams contributing to the one-loop gluon self-energy.}
\end{figure}

In the following, we compute the real-time gluon self-energies at one-loop order where four Feynman diagrams as shown in Fig.~\ref{gse} contribute.\footnote{Although we adopt the double line notation in this work, for simplicity, the Feynman diagrams are drawn in the usual manner, i.e., the gluon and ghost lines are not doubled.} We are interested in the physical ``11" component of the gluon self-energy which can be written as $\Pi_{11}=(\Pi_R+\Pi_A+\Pi_F)/2$. For convenience, the calculation of the retarded/advanced and symmetric gluon self-energies will be carried out separately.

\section{The retarded/advanced gluon self-energy in a semi-quark-gluon plasma}
\la{pire}

We start by considering the quark-loop diagram for the retarded gluon self-energy. Using the Feynman rules as provided in Sec.~\ref{feynman}, we can show
\ba\la{pir}
\Pi^{ab,cd}_{R;\mu \nu}(P,\cQ)&=&\Pi^{ab,cd}_{11;\mu \nu}(P,\cQ)+\Pi^{ab,cd}_{12;\mu \nu}(P,\cQ)= i g^2 N_f \sum_{ef}(t^{ab})_{ef}(t^{cd})_{fe}\int\frac{d^4 K}{(2\pi)^4} {\rm Tr}[\gamma_\mu \displaystyle{\not} Q \gamma_\nu  \displaystyle{\not} K]\,\nonumber \\
&\times&\big[\tilde{D}_{11}(Q,\cQ^f)\tilde{D}_{11}(K,\cQ^e)-\tilde{D}_{21}(Q,\cQ^f)\tilde{D}_{12}(K,\cQ^e)\big]\,.
\ea
In the above equation, $K=P+Q$ and an overall factor $(-1)$ coming from the fermion loop has been included. In addition, the minus sign in the square bracket appears due to the type-$2$ field. Performing the trace over the $\gamma$-matrices and expressing the bare propagators in terms of the three independent components in the Keldysh representation, we arrive at
\ba\la{pir2}
\Pi^{ab,cd}_{R;\mu \nu}(P,\cQ) &=& i g^2 N_f \sum_{ef}{\cal P}^{ab,fe}{\cal P}^{cd,ef} \int\frac{d^4 K}{(2\pi)^4}  (2 K_\mu K_\nu-P_\mu K_\nu-K_\mu P_\nu-g_{\mu\nu} Q \cdot K)\nonumber \\
&\times& \big[\tilde{D}_{F}(K,\cQ^e)\tilde{D}_{A}(Q)+\tilde{D}_{R}(K)\tilde{D}_{F}(Q,\cQ^f)\big] \,.
\ea
Notice that terms proportional to $\tilde{D}_{R}(K)\tilde{D}_{R}(Q)$ or $\tilde{D}_{A}(K)\tilde{D}_{A}(Q)$ are independent of the background field and dropped in the above equation because their contributions vanish after integrating over $k_0$.

Furthermore, for vanishing background field, the two terms, $\sim\tilde{D}_{F}(K)\tilde{D}_{A}(Q)$ and $\sim\tilde{D}_{R}(K)\tilde{D}_{F}(Q)$, contribute equally to the retarded gluon self-energy. Although this is no longer true in the semi-QGP, there exists a simple relation between the two terms as given by the following equation:
\ba\la{re}
&&\int\frac{d^4 K}{(2\pi)^4}  (2 K_\mu K_\nu-P_\mu K_\nu-K_\mu P_\nu-g_{\mu\nu} Q \cdot K)\tilde{D}_{R}(K)\tilde{D}_{F}(Q,\cQ^f)\nonumber \\
&=&\int\frac{d^4 Q}{(2\pi)^4}  \big[2 Q_\mu Q_\nu+P_\mu Q_\nu+Q_\mu P_\nu-g_{\mu\nu} Q \cdot (P+Q)\big]\tilde{D}_{R}(P+Q)\tilde{D}_{F}(Q,\cQ^f)\nonumber \\
&=&\int\frac{d^4 K}{(2\pi)^4}  \big[2 K_\mu K_\nu-P_\mu K_\nu-K_\mu P_\nu-g_{\mu\nu} K \cdot (K-P)\big]\tilde{D}_{R}(P-K)\tilde{D}_{F}(-K,\cQ^f)\nonumber \\
&=&\int\frac{d^4 K}{(2\pi)^4}  \big[2 K_\mu K_\nu-P_\mu K_\nu-K_\mu P_\nu-g_{\mu\nu} K \cdot (K-P)\big]\tilde{D}_{A}(K-P)\tilde{D}_{F}(K,-\cQ^f)\,. \nonumber \\
\ea
In the third line of the above equation, the variable $Q$ is changed into $-K$. In addition, we have used $\tilde{D}_{R}(P-K)=\tilde{D}_{A}(K-P)$ and replaced $\tilde{D}_{F}(-K,\cQ^f)$ with $\tilde{D}_{F}(K,-\cQ^f)$ which is valid under the integration $\int d k_0$ with the delta function $\delta(K^2)$. 

To proceed further, we should make use of the HTL approximation. Taking the spatial components as an example, we will encounter the following integral:
\ba
&& i \int\frac{d^4 K}{(2\pi)^4}  \big[2 k_i k_j-p_i k_j-k_i p_j-g_{ij} (K-P) \cdot K\big]\tilde{D}_{F}(K,\cQ^e)\tilde{D}_{A}(Q)\nonumber \\
&=& \int\frac{d^3 {\bf k}}{(2\pi)^3} \frac{1}{k}\Big\{[2 k_i k_j-p_i k_j-k_i p_j+g_{ij} (p_0k-{\bf k}\cdot{\bf p})]\frac{ {\tilde n}_+^e(k)}{-2p_0k+2{\bf k}\cdot {\bf p}+P^2-i \epsilon} \nonumber\\
&&+[2 k_i k_j-p_i k_j-k_i p_j+g_{ij} (-p_0k-{\bf k}\cdot {\bf p})]\frac{ {\tilde n}_-^e(k)}{2p_0k+2{\bf k}\cdot{\bf p}+P^2+i \epsilon}\Big\}\, .
\ea
Changing the integral variable ${\bf k}\rightarrow -{\bf k}$, the leading order (LO) contribution can be written as the following compact form
\ba
&&  \int\frac{d^3 {\bf k}}{(2\pi)^3} \frac{1}{k} \frac{2 k_i k_j}{-2p_0k+2{\bf k}\cdot {\bf p}-i \epsilon}\big({\tilde n}_+^e(k)-{\tilde n}_-^e(k)\big)\nonumber \\
&=&  \frac{1}{p_0}  \int \frac{dk}{2\pi^2} k^2 \big({\tilde n}_+^e(k)-{\tilde n}_-^e(k)\big) \int \frac{d \Omega}{4\pi} p_0 \frac{ {\hat k}_i {\hat k}_j}{- p_0+{\hat {\bf k}}\cdot {\bf p}-i \epsilon} \, ,
\ea
where ${\hat {\bf k}}$ is defined as ${\bf k}/k$. We should mention that the LO contribution vanishes in the limit of  zero background field because the two distribution functions ${\tilde n}_+^e(k)$ and ${\tilde n}_-^e(k)$ become identical. In semi-QGP with a nonvanishing background field, integrating over $k$ leads to a contribution $\sim i T^3 B_3({\tilde {\sf q}}^e)/p_0$ where the third Bernoulli polynomial $B_3(x)$ is given by Eq.~(\ref{aa3}). Here, we also define the dimensionless background field ${\sf q}=\cQ/(2\pi T)$ and ${\tilde {\sf q}}^e={\sf q}^e+1/2=\cQ ^e/(2\pi T)+1/2$.  

The NLO contribution can be obtained as the following:
\ba
&&\int\frac{d^3 {\bf k}}{(2\pi)^3}\frac{{\tilde n}_+^e(k)+{\tilde n}_-^e(k)}{k} \Big[\frac{- p_i k_j-p_j k_i +g_{ij}(p_0k-{\bf k}\cdot {\bf p})}{-2p_0k+2{\bf k}\cdot {\bf p}-i \epsilon}-\frac{2k_i k_j P^2}{(-2p_0k+2{\bf k}\cdot {\bf p}-i \epsilon)^2}\Big]\nonumber \\
&=&  \int \frac{d^3 {\bf k}}{(2\pi)^3} \frac{1}{2}\big({\tilde n}_+^e(k)+{\tilde n}_-^e(k)\big) \Big[p_l \partial\Big(\frac{-{\hat k}_i{\hat k}_j}{-p_0+{\hat {\bf k}}\cdot {\bf p}-i \epsilon}\Big)/\partial k_l+\frac{\partial {\hat k}_i}{\partial k_j}\Big] \nonumber \\
&=& \int \frac{dk}{4\pi^2} k^2  \frac{\partial\big({\tilde n}_+^e(k)+{\tilde n}_-^e(k)\big)}{\partial k} \int \frac{d \Omega}{4\pi} p_0\frac{ {\hat k}_i {\hat k}_j}{-p_0+{\hat {\bf k}}\cdot {\bf p}-i \epsilon}\nonumber \\
&=& - \int \frac{dk}{2\pi^2} k \big({\tilde n}_+^e(k)+{\tilde n}_-^e(k)\big) \int \frac{d \Omega}{4\pi} p_0\frac{ {\hat k}_i {\hat k}_j}{-p_0+{\hat {\bf k}}\cdot {\bf p}-i \epsilon} \, ,
\ea
where we have integrated by part to get the final expression. Technically, the above calculation does not involve anything new as compared to the vanishing background case. After integrating over $k$, we find that instead of a contribution $\sim T^2$ at vanishing background field, when $\cQ\neq 0$, the above equation gives rise to a contribution $\sim T^2 B_2({\tilde {\sf q}}^e)$ where modifications due to the nonzero background field are entirely encoded in the second Bernoulli polynomial $B_2(x)$.

The other Lorentz components can be computed in a similar way and the result up to NLO reads
\ba\la{term1}
&& i \int\frac{d^4 K}{(2\pi)^4}  \big[2 K_\mu K_\nu-P_\mu K_\nu-K_\mu P_\nu-g_{\mu\nu} (K-P) \cdot K\big]\tilde{D}_{F}(K,\cQ^e)\tilde{D}_{A}(Q)\nonumber \\
&=&- \frac{1}{p_0} \int \frac{dk}{2\pi^2} k^2  \big({\tilde n}_+^e(k)-{\tilde n}_-^e(k)\big) \Gamma^{(1)}_{\mu\nu}(P)- \int \frac{dk}{2\pi^2} k \big({\tilde n}_+^e(k)+{\tilde n}_-^e(k)\big)\Gamma^{(2)}_{\mu\nu}(P) \,,
\ea
where the integral over the solid angle leads to the following dimensionless functions
\ba\la{insa}
\Gamma ^{(1)}_{\mu\nu}(P) &=& \int \frac{d \Omega}{4\pi} p_0 \frac{ {\hat K}_\mu {\hat K}_\nu}{ P\cdot {\hat K}+i \epsilon}\, ,\nonumber \\
\Gamma^{(2)}_{\mu\nu}(P)&= &\int \frac{d \Omega}{4\pi} \Big( M_\mu M_\nu-p_0 \frac{ {\hat K}_\mu {\hat K}_\nu}{P\cdot {\hat K}+i \epsilon}\Big)\, ,
\ea
where ${\hat K}_\mu=(1,-\hat{ \bf{k}})$ and $M_{\mu}$ is the heat-bath vector, which in the local rest frame is given by $M_\mu = (1, 0, 0, 0)$. These two functions have nothing to do with the background field and satisfy
\be
\Gamma^{(1)}_{\mu\nu}(P)=M_\mu M_\nu-\Gamma^{(2)}_{\mu\nu}(P)\, .
\ee
In terms of the mutually orthogonal projection operators $A_{\mu \nu}(P)$ and $B_{\mu \nu}(P)$, one can express $\Gamma^{(2)}_{\mu\nu}(P)$ as
\be\label{defg2}
\Gamma^{(2)}_{\mu\nu}(P)=\Pi_T(P) A_{\mu\nu}(P)+\Pi_L(P) B_{\mu\nu}(P)\, ,
\ee
where longitudinal and transverse structure functions are given by
\ba\la{pitl}
\Pi_T(P)&=&\frac{p_0^2}{2p^2}\Big(\frac{p_0^2-p^2}{2p_0p} \ln\frac{p_0+p+ i\e}{p_0-p+i \e}-1\Big)\, ,\nonumber \\
\Pi_L(P)&=&\frac{p_0^2}{p^2}\Big(1-\frac{p_0}{2p} \ln\frac{p_0+p+ i\e}{p_0-p+i \e}\Big)\, ,
\ea
and the two projection operators read
\ba\la{po}
A_{\mu\nu}(P)&=&- g_{\mu\nu} + \frac{P_\mu P_\nu}{P^2}+\frac{{\tilde M}_\mu {\tilde M}_\nu} {{\tilde M}^2}\, ,\nonumber \\
B_{\mu\nu}(P)&=&- \frac{P^2}{(M\cdot P)^2} \frac{{\tilde M}_\mu {\tilde M}_\nu}{{\tilde M}^2}\, .
\ea
In the above equation, ${\tilde M}_\mu$ is orthogonal to $P_\mu$ which is defined as
\be
{\tilde M}_\mu= M_\mu-\frac{M\cdot P}{P^2} P_\mu\, .
\ee

According to Eq.~(\ref{re}),  the other contribution in the retarded gluon self-energy can be obtained by changing ${\cQ}^e$ into $-{\cQ}^f$ in Eq.~(\ref{term1}). Then, we arrive at
\ba\la{pirenew}
&&\Pi^{ab,cd}_{R;\mu \nu}(P,\cQ) =- g^2 N_f \sum_{ef}\bigg\{\frac{1}{p_0}\int \frac{dk}{2\pi^2} k^2  \big[({\tilde n}_+^e(k)-{\tilde n}_-^e(k))-({\tilde n}_+^f(k)-{\tilde n}_-^f(k))\big] \Gamma^{(1)}_{\mu\nu}(P) \nonumber \\
&&+  \int \frac{dk}{2\pi^2} k \big[ ({\tilde n}_+^e(k)+{\tilde n}_-^e(k)) +({\tilde n}_+^f(k)+{\tilde n}_-^f(k))\big]\Gamma^{(2)}_{\mu\nu}(P)\bigg\} {\cal P}^{ab,fe}{\cal P}^{cd,ef}\, .
\ea


At $\cQ=0$, the gluon self-energy is simply proportional to an identity matrix in the color space.\footnote{When the standard choice for the generators of a gauge group is adopted, the identity matrix is $\delta^{AB}$ where $A$ and $B$ refer to adjoint indices running from $1$ to $N^2-1$ for $SU(N)$. With the double line notation, the identity matrix is given by ${\cal P}^{ab,cd}$.} Switching on the background field, on the other hand, the color structure becomes nontrivial as indicated by the following equation:
\ba\la{cs}
&&\sum_{ef}{\cal P}^{ab}_{ef}{\cal P}^{cd}_{fe}=\Big(\delta^{ae}\delta^{bf}-\frac{1}{N}\delta^{ab}\delta^{ef}\Big)\Big(\delta^{cf}\delta^{de}-\frac{1}{N}\delta^{cd}\delta^{ef}\Big)\nonumber\\
&&\quad=\delta^{ad}\delta^{bc}\Big|_{e=a,f=c}-\frac{1}{N}\delta^{ab}\delta^{cd}\Big|_{e=f=a}-\frac{1}{N}\delta^{ab}\delta^{cd}\Big|_{e=f=c}+\frac{1}{N^2}\delta^{ab}\delta^{cd}\sum_{e}\Big|_{f=e}\, .
\ea
For the off-diagonal components $\sim\delta^{ad}\delta^{bc}$, the color indices $e$ and $f$ in the Bernoulli polynomials in Eq.~(\ref{pirenew}) are replaced by $a$ and $c$, respectively. On the other hand, by requiring $e=f$, there are no diagonal components $\sim \delta^{ab}\delta^{cd}$ existing in the LO contribution. Explicitly, the retarded gluon self-energy takes the following form:
\ba\la{pirlo}
\Pi^{ab,cd}_{R;\mu \nu}  \big|_{{\rm LO}}(P,\cQ) &= &- \frac{g^2}{2\pi^2 p_0} N_f \delta^{ad}\delta^{bc}  \int k^2 dk  \big[({\tilde n}_+^a(k)-{\tilde n}_-^a(k))-({\tilde n}_+^c(k)-{\tilde n}_-^c(k))\big] \Gamma^{(1)}_{\mu\nu}(P)\, \nonumber\\
&=& i \frac{T}{p_0} \frac{N_f}{ 6} g^2 T^2 \delta^{ad}\delta^{bc}{\cal G}_f({\sf q}^a,{\sf q}^c) \Gamma^{(1)}_{\mu\nu}(P)\, ,
\ea
at LO in the HTL approximation and the corresponding result at NLO is given by
\ba\la{pirnlo}
&&\Pi^{ab,cd}_{R;\mu \nu}  \big|_{{\rm NLO}}(P,\cQ)=\frac{g^2}{2\pi^2} N_f \int k d k \Big\{ \delta^{ab}\delta^{cd}  \frac{2}{N}  \Big[\big({\tilde n}_+^a(k)+{\tilde n}_-^a(k)\big)+ \big({\tilde n}_+^c(k)+{\tilde n}_-^c(k)\big)\,\nonumber\\
&&\quad-\frac{1}{N}\sum_e \big({\tilde n}_+^e(k)+{\tilde n}_-^e(k)\big)\Big]  - \delta^{ad}\delta^{bc}  \Big[ \big({\tilde n}_+^a(k)+{\tilde n}_-^a(k)\big) +\big({\tilde n}_+^c(k)+{\tilde n}_-^c(k)\big)\Big] \Big\} \Gamma^{(2)}_{\mu\nu}(P)\, \nonumber \\
&&\quad= \frac{N_f}{6} g^2 T^2\Big[ \delta^{ab}\delta^{cd}  \frac{1}{N}{\cal F}^{(2)}_f({\sf q}^a,{\sf q}^c) - \delta^{ad}\delta^{bc} {\cal F}^{(1)}_f({\sf q}^a,{\sf q}^c)\Big]\Gamma^{(2)}_{\mu\nu}\,.
\ea
In the above equations, we have used the following fermionic integrals:
\ba
\int k dk ({\tilde n}_+^{a}(k)+{\tilde n}_-^{a}(k))&=&-2\pi^2T^2 B_2({\tilde{\sf q}}^a)\, ,\nonumber\\
\int k^2 dk ({\tilde n}_+^{a}(k)-{\tilde n}_-^{a}(k))&=&- i \frac{8 \pi^3T^3}{3} B_3({\tilde{\sf q}}^a)\,,
\ea
and the $\cQ$-dependent functions are defined by
\ba\la{qdf}
{\cal G}_f({\sf q}^a,{\sf q}^c)&=& 8\pi  (B_3({\tilde{\sf q}}^a)-B_3({\tilde{\sf q}}^c))\, , \quad {\cal F}^{(1)}_f({\sf q}^a,{\sf q}^c)=- 6 (B_2({\tilde{\sf q}}^a)+B_2({\tilde{\sf q}}^c))\, , \nonumber \\
{\cal F}^{(2)}_f({\sf q}^a,{\sf q}^c)&=& \frac{12}{N}\sum_{e}B_2({\tilde{\sf q}}^e)-12(B_2({\tilde{\sf q}}^a)+B_2({\tilde{\sf q}}^c))\, .
\ea

The three Feynman diagrams from the pure gauge part in Fig.~\ref{gse} lead to the following contribution which takes a form analogous to Eq.~(\ref{pir2})\footnote{Some details of the calculation in the double line basis can be found in Appendix~\ref{app}.},
\ba\la{pirglu}
\Pi^{ab,cd}_{R;\mu \nu}(P,\cQ)& =& i g^2\sum_{efgh} f^{ab,fe,gh} f^{cd,hg,ef} \int\frac{d^4 K}{(2\pi)^4}  (2 K_\mu K_\nu-P_\mu K_\nu-K_\mu P_\nu-g_{\mu\nu} Q \cdot K)\,\nonumber \\
&\times&[D_{F}(K,\cQ^{ef})D_{A}(Q)+D_{R}(K)D_{F}(Q,\cQ^{gh})]\,.
\ea
As compared to Eq.~(\ref{pir2}), the only nontrivial difference lies in the color structures. In addition, following the same analysis as given in Eq.~(\ref{re}), it can be shown that the two terms in the square bracket give a similar contribution to the retarded gluon self-energy, in other words, $D_{R}(K)D_{F}(Q,\cQ^{gh})$ can be replaced by $D_{F}(K,\cQ^{hg})D_{A}(Q)$ in the above equation.

Summing over the color indices in Eq.~(\ref{pirglu}), we obtain
\ba\la{cs2}
&&\sum_{efgh} f^{ab,fe,gh} f^{cd,hg,ef} = - \frac{1}{2}\Big(\delta^{ae}\delta^{fh}\delta^{bg}-\delta^{ah}\delta^{bf}\delta^{eg}\Big)\Big(\delta^{cg}\delta^{hf}\delta^{de}-\delta^{cf}\delta^{dh}\delta^{ge}\Big)\nonumber\\
&&\quad=\frac{1}{2}\Big(\delta^{ab}\delta^{cd}\Big|_{\substack{ef=ac\\gh=ac}}+\delta^{ab}\delta^{cd}\Big|_{\substack{ef=ca\\gh=ca}}-\delta^{ad}\delta^{bc}\sum_e\Big|_{\substack{f=c\\gh=ea}}-\delta^{ad}\delta^{bc}\sum_f\Big|_{\substack{e=a\\gh=cf}}\Big)\, .
\ea
The integral over the hard momentum $K$ in Eq.~(\ref{pirglu}) can be carried out with the same procedures as used in the calculation of the quark-loop diagram. Therefore, it is straightforward to obtain the contributions from the pure gauge part which are given by
\be\la{pirlog}
\Pi^{ab,cd}_{R;\mu \nu}  \big|_{{\rm LO}}(P,\cQ) =- i \frac{T}{p_0} \frac{N}{3} g^2 T^2 \delta^{ad}\delta^{bc} {\cal G}_g({\sf q}^a,{\sf q}^c) \Gamma^{(1)}_{\mu\nu}(P)\, ,
\ee
at LO, while the NLO result reads
\be\la{pirnlog}
\Pi^{ab,cd}_{R;\mu \nu}  \big|_{{\rm NLO}}(P,\cQ)= \frac{N}{3}g^2 T^2 \Big[\delta^{ab}\delta^{cd}\frac{1}{N} {\cal F}^{(2)}_g({\sf q}^a,{\sf q}^c)  -\delta^{ad}\delta^{bc} {\cal F}^{(1)}_g({\sf q}^a,{\sf q}^c) \Big]\Gamma^{(2)}_{\mu\nu}(P)\, .
\ee
Here, ${\sf q}^{ac}={\sf q}^{a}-{\sf q}^{c}=(\cQ ^a-\cQ ^c)/(2\pi T)$ and the following bosonic integrals are used,
\ba
\int k dk (n_+^{ac}(k)+n_-^{ac}(k))&=&2\pi^2T^2 B_2({\sf q}^{ac})\, ,\nonumber\\
\int k^2 dk (n_+^{ac}(k)-n_-^{ac}(k))&=& i \frac{8 \pi^3T^3}{3} B_3({\sf q}^{ac})\,.
\ea
Similar as before, we also define the $\cQ$-dependent functions for the pure gauge part,
\ba\la{qdfpg}
{\cal G}_g({\sf q}^a,{\sf q}^c) =\frac{4\pi}{N} \sum_e(B_3({\sf q}^{ae})+B_3({\sf q}^{ec}))\, , &&\quad {\cal F}^{(1)}_g({\sf q}^a,{\sf q}^c)= \frac{3}{N} \sum_e(B_2({\sf q}^{ae})+B_2({\sf q}^{ec})) \, , \nonumber \\
{\cal F}^{(2)}_g({\sf q}^a,{\sf q}^c) &=& 6B_2({\sf q}^{ac})\, .
\ea

Summing up the above results, the final expression for the retarded gluon self-energy $\Pi^{ab,cd}_{R;\mu \nu}(P,\cQ)$ can be obtained as
\be\la{pirtotlo}
\Pi^{ab,cd}_{R;\mu \nu}  \big|_{{\rm LO}}(P,\cQ)=\Pi^{\rm ano}_{R;\mu \nu}(P,\cQ)= i \frac{T}{p_0}   \delta^{ad} \delta^{bc} \big[m_f^2 {\cal G}_f({\sf q}^a,{\sf q}^c)- m_g^2 {\cal G}_g({\sf q}^a,{\sf q}^c) \big] \Gamma^{(1)}_{\mu\nu}(P)\, ,
\ee
and
\ba
\la{pirtotnlo}
\Pi^{ab,cd}_{R;\mu \nu}  \big|_{{\rm NLO}}(P,\cQ)&=&\Pi^{\rm nor}_{R;\mu \nu}(P,\cQ)= -  \bigg\{ m_f^2 \Big[\delta^{ad} \delta^{bc}  {\cal F}^{(1)}_f({\sf q}^a,{\sf q}^c)-\frac{1}{N}\delta^{ab} \delta^{cd} {\cal F}^{(2)}_f({\sf q}^a,{\sf q}^c)\Big]\,\nonumber\\
&+&m_g^2 \Big[\delta^{ad} \delta^{bc} {\cal F}^{(1)}_g({\sf q}^a,{\sf q}^c)-\frac{1}{N}\delta^{ab} \delta^{cd} {\cal F}^{(2)}_g({\sf q}^a,{\sf q}^c)\Big]\bigg\} \Gamma^{(2)}_{\mu\nu}(P) \, ,
\ea
where $m_f^2$ and $m_g^2$ denote the fermionic and bosonic contributions in the Debye mass square, respectively. By definition, $m_D^2=m_f^2+m_g^2$ with $m_f^2=g^2T^2 N_f/6$ and $m_g^2=g^2T^2 N/3$.

Following the terminology in Ref.~\cite{Guo:2020jvc}, we introduce $\Pi^{\rm ano}_{R;\mu \nu}(P,\cQ)$ in Eq.~(\ref{pirtotlo}) to denote the anomalous contributions in the retarded gluon self-energy. These contributions arise at LO in the HTL approximation, $\sim g^2T^3/p_0$, and are not transverse since $P^\mu \Gamma^{(1)}_{\mu\nu}(P) \neq 0$. According to Eq.~(\ref{pirlo}), the integral over $k$ involves a difference between two distribution functions $(n_+^e(k)-n_-^e(k))$, therefore, the anomalous contributions only show up at $\cQ\neq 0$. In addition, Eq.~(\ref{pirtotlo}) is antisymmetric under the interchange of the color indices $a\leftrightarrow c$, we can easily show the following identity,
\be\la{idsum}
\sum_{abcd}{\cal P}^{ab,cd}\Pi^{ab,cd}_{R;\mu \nu}  \big|_{{\rm LO}}(P,\cQ)=0\, .
\ee

Accordingly, terms associated with $(n_+^e(k)+n_-^e(k))$ lead to the normal contributions $\Pi^{\rm nor}_{R;\mu \nu}(P,\cQ)$ in Eq.~(\ref{pirtotnlo}). They emerge at NLO, $\sim g^2 T^2$, where the same dimensionless function $\Gamma^{(2)}_{\mu\nu}(P)$ as at $\cQ=0$ is now multiplied by a $\cQ$-dependent mass squared defined as
\ba\la{qmd}
({\cal M}_D^2)^{ab,cd}(\cQ)&=&m_f^2 \Big[\delta^{ad} \delta^{bc}  {\cal F}^{(1)}_f({\sf q}^a,{\sf q}^c)-\frac{1}{N}\delta^{ab} \delta^{cd} {\cal F}^{(2)}_f({\sf q}^a,{\sf q}^c)\Big]\,\nonumber \\
&+&m_g^2 \Big[\delta^{ad} \delta^{bc} {\cal F}^{(1)}_g({\sf q}^a,{\sf q}^c)-\frac{1}{N}\delta^{ab} \delta^{cd} {\cal F}^{(2)}_g({\sf q}^a,{\sf q}^c)\Big]\, .
\ea
As we can see, modifications on the Debye mass due to the nonzero background field can be entirely encoded in the second Bernoulli polynomials $B_2(x)$. Then, we can express the normal contributions as
\be\la{effpir}
\Pi^{\rm nor}_{R;\mu \nu}(P,\cQ)=- ({\cal M}_D^2)^{ab,cd}(\cQ) \Gamma^{(2)}_{\mu\nu}(P) \, .
\ee
Obviously, the above gluon self-energy is transverse because the dimensionless function $\Gamma^{(2)}_{\mu\nu}(P)$ is orthogonal to $P^\mu$.

Considering vanishing background field, we find ${\cal G}_f({\sf q}^a,{\sf q}^c)={\cal G}_g({\sf q}^a,{\sf q}^c)=0$ and the other four $\cQ$-dependent functions in Eqs.~(\ref{qdf}) and (\ref{qdfpg}) equal one. Therefore, the retarded gluon self-energy reduces to the following well-known result as expected,
\be\la{pir0}
\Pi^{ab,cd}_{R;\mu \nu}(P,\cQ\rightarrow 0)= - m_D^2 \Gamma^{(2)}_{\mu\nu}(P){\cal P}^{ab,cd}\, .
\ee

Given the above discussions, the calculation of the advanced gluon self-energy is trivial. As compared to the retarded one, the only difference is that the $+ i \epsilon$ description is now replaced by $- i \epsilon$ in Eq.~(\ref{pitl}). In addition, these results can be also obtained from the imaginary time gluon self-energy as obtained in Ref.~\cite{Hidaka:2009hs} after an analytical continuation $i \omega_n + i \cQ^{ab} \rightarrow p_0 \pm i \epsilon$ where $\omega_n$ is the bosonic Matsubara frequency.

\section{The symmetric gluon self-energy in a semi-quark-gluon plasma}
\la{pisy}

It is also interesting to study the symmetric gluon self-energy in a semi-quark-gluon plasma which has not been addressed in previous studies. As is well known, at vanishing background field, the three independent gluon self-energies in the Keldysh representation satisfy the KMS condition as given by Eq.~(\ref{kmseq}). Therefore, the LO contribution of $\Pi^{ab,cd}_{F;\mu \nu}(P,\cQ= 0)$ can be simply obtained from $\Pi^{ab,cd}_{R/A;\mu \nu}(P,\cQ= 0)$ at NLO  (the LO contribution vanishes as discussed in the previous section) where the HTL approximation should also be imposed on the distribution function, leading to an extra $\sim T/p_0$ enhancement. However, in the presence of a nonzero background field, the KMS condition does not appear to be a trivial extension of that at $\cQ=0$ because one has to incorporate some new terms when $\cQ \neq 0$, such as the nonzero LO contributions in the retarded/advanced gluon self-energy. In the following, we present the explicit calculation for $\Pi^{ab,cd}_{F;\mu \nu}(P,\cQ)$ up to NLO in the HTL approximation which is the same order as we compute for the retarded and advanced ones. As we will see, such a calculation is necessary to understand the KMS condition in a semi-QGP with $\cQ\neq 0$.

\subsection{The leading order contributions in the HTL approximation}

Summing the ``11" and ``22" components, the contribution from the quark-loop diagram can be obtained as
\ba\la{pif}
\Pi^{ab,cd}_{F;\mu \nu}(P,\cQ) &=&\Pi^{ab,cd}_{11;\mu \nu}(P,\cQ)+\Pi^{ab,cd}_{22;\mu \nu}(P,\cQ)= i g^2 N_f \sum_{ef} (t^{ab})_{ef}(t^{cd})_{fe}\int\frac{d^4 K}{(2\pi)^4}  {\rm Tr}[\gamma_\mu \displaystyle{\not} Q \gamma_\nu  \displaystyle{\not} K]\, \nonumber \\
&\times&\big[\tilde{D}_{11}(Q,\cQ^f)\tilde{D}_{11}(K,\cQ^e)+\tilde{D}_{22}(Q,\cQ^f)\tilde{D}_{22}(K,\cQ^e)\big]\, \nonumber \\
&=& i g^2 N_f \sum_{ef}{\cal P}^{ab,fe}{\cal P}^{cd,ef} \int\frac{d^4 K}{(2\pi)^4}  (2 K_\mu K_\nu-P_\mu K_\nu-K_\mu P_\nu-g_{\mu\nu} Q \cdot K) \nonumber \\
&\times& \big[{\tilde D}_{F}(Q,\cQ^{f}) {\tilde D}_{F}(K,\cQ^{e})-\big( {\tilde D}_{R}(K)- {\tilde D}_{A}(K)\big)\big( {\tilde D}_{R}(Q)-{\tilde D}_{A}(Q)\big)\big]\nonumber \\
&=& i g^2 N_f \sum_{ef}{\cal P}^{ab,fe}{\cal P}^{cd,ef} \int\frac{d^4 K}{(2\pi)^4}  (2 K_\mu K_\nu-P_\mu K_\nu-K_\mu P_\nu-g_{\mu\nu} Q \cdot K) \nonumber \\
&\times& 4\pi^2 \delta(K^2)\delta(Q^2) \big[(1-2 {\tilde n}(q_0,\cQ^f))(1-2 {\tilde n}(k_0,\cQ^e))-{\rm sgn}(q_0){\rm sgn}(k_0)\big]\,,
\ea
where we used $\tilde{D}_R(K)-\tilde{D}_A(K)=2\pi \delta(K^2)\, {\rm sgn}(k_0)$. After performing the integral over $k_0$ and keeping only the LO contributions in the HTL approximation, the symmetric gluon self-energy can be written as
\ba\la{pifij}
\Pi^{ab,cd}_{F;ij}(P,\cQ) &=& i 2 \pi g^2 N_f \sum_{ef} \int \frac{d^3 {\bf k}}{(2\pi)^3} \frac{ {\hat k}_i  {\hat k}_j}{ p}\Big[ \big(2 {\tilde n}_+^f(k) {\tilde n}_+^e(k)- {\tilde n}_+^f(k)- {\tilde n}_+^e(k)\big)\delta({\hat {\bf k}}\cdot {\hat {\bf p}}-p_0/p)\, \nonumber \\
&+& \big(2 {\tilde n}_-^f(k) {\tilde n}_-^e(k)-{\tilde n}_-^f(k)- {\tilde n}_-^e(k)\big)\delta({\hat {\bf k}}\cdot {\hat {\bf p}}+p_0/p)\Big]{\cal P}^{ab,fe}{\cal P}^{cd,ef}\, .
\ea
To make the notations compact in the above equation, we take the spatial components as an example. As before, the corresponding integral over the solid angle is independent of the background field which leads to
\be\la{inso}
\int d \Omega {\hat k}_i  {\hat k}_j \delta({\hat {\bf k}}\cdot {\hat {\bf p}}\pm p_0/p) =\Pi^\prime_T(P) \Big(\delta_{ij}-\frac{p_i p_j}{p^2}\Big)+\Pi^\prime_L(P) \frac{p_i p_j}{p^2}\, ,
\ee
where the two structure functions take simpler forms as compared to Eq.~(\ref{pitl}),
\be
\Pi^\prime_T(P)= \pi \frac{p^2-p_0^2}{p^2} \theta(p^2-p_0^2)\,\quad\quad {\rm and}\quad\quad \Pi^\prime_L(P)= 2 \pi \frac{p_0^2}{p^2} \theta(p^2-p_0^2)\,.
\ee
The other Lorentz components can be computed in a similar way and we can show that
\be\la{pif3}
\Pi^{ab,cd}_{F;\mu\nu}(P,\cQ) = i  \frac{g^2}{4\pi^2 p} N_f \sum_{ef} {\cal P}^{ab,fe}{\cal P}^{cd,ef}\int k^2 dk \sum_{\sigma=\pm} \big[2 {\tilde n}_\sigma^f(k) {\tilde n}_\sigma^e(k)- {\tilde n}_\sigma^f(k)- {\tilde n}_\sigma^e(k)\big] \Lambda_{\mu\nu}(P)\, .
\ee
In the above equation, the dimensionless function $\Lambda_{\mu\nu}(P)$ is given by
\be
\Lambda_{\mu\nu}(P)\equiv\Pi^\prime_T(P) A_{\mu\nu}(P)+\Pi^\prime_L(P) B_{\mu\nu}(P)\, ,
\ee
and the two projection operators $A_{\mu \nu}(P)$ and $B_{\mu \nu}(P)$ are the same as before which have been defined in Eq.~(\ref{po}).

Equation~(\ref{pif3}) indicates that nonzero background field only modifies the integral over $k$ as the distribution functions are $\cQ$-dependent now. Performing the integral in Eq.~(\ref{pif3}), we arrive at
\be\label{piflofer}
\Pi^{ab,cd}_{F;\mu \nu}\big|_{{\rm LO}}(P,\cQ) =  i \frac{T}{p} m_f^2\Big[\frac{1}{2}  \delta^{ad} \delta^{bc} \cot(\pi {\sf q}^{ac}) {\cal G}_f({\sf q}^a,{\sf q}^c)+\frac{1}{N} \delta^{ab} \delta^{cd} {\cal F}^{(2)}_f({\sf q}^a,{\sf q}^c)\Big] \Lambda_{\mu\nu}(P)\, .
\ee
As compared to $\Pi^{ab,cd}_{R;\mu \nu}\big|_{{\rm LO}}(P,\cQ) $, the most important difference lies in the fact that the LO terms in the HTL approximation have a nonvanishing contribution to $\Pi^{ab,cd}_{F;\mu\nu} (P,\cQ)$ even at $\cQ=0$. In the above result, to avoid an ambiguous expression of the type ``$\infty \cdot0$ " which originates from $\sim \cot(\pi {\sf q}^{ac}){\cal G}_{f}({\sf q}^a,{\sf q}^c)$, the special case ${\sf q}^a={\sf q}^c$ should be understood as ${\sf q}^a\rightarrow {\sf q}^c \pm \epsilon$. Consequently, one can show that
\be
B_3({\tilde{\sf q}}^a)-B_3({\tilde{\sf q}}^c)\rightarrow \pm 3 B_2({\tilde{\sf q}}^c) \epsilon \quad {\rm and}\quad \cot(\pi {\sf q}^{ac}) \rightarrow \pm 1/(\pi \epsilon)\, .
\ee
Therefore,  in this limit of $\cQ\rightarrow 0$, the square bracket in Eq.~(\ref{piflofer}) reduces to $-{\cal P}^{ab,cd}$.

The LO contributions from the pure gauge part to $\Pi^{ab,cd}_{F;\mu \nu}(P,\cQ)$ are similar to those from the quark-loop diagram. In particular, one only needs to do the following replacements in Eq.~(\ref{pifij}),
\ba
N_f \sum_{ef} {\cal P}^{ab,fe}{\cal P}^{cd,ef} &\rightarrow& \sum_{efgh} f^{ab,fe,gh} f^{cd,hg,ef}\, ,\nonumber \\
{\tilde n}_\pm^e(k) \rightarrow -n_{\pm}^{ef}(k)\,,\quad &{\rm and}& \quad {\tilde n}_\pm^f(k) \rightarrow -n_{\pm}^{gh}(k)\, .
\ea
By using a set of bosonic integrals, the results of the pure gauge part can be simply obtained as the following:
\be\label{piflobo}
\Pi^{ab,cd}_{F;\mu \nu}\big|_{{\rm LO}}(P,\cQ)=  i \frac{T}{p}
m_g^2\Big[-\frac{1}{2}  \delta^{ad} \delta^{bc} \cot(\pi {\sf q}^{ac}) {\cal G}_g({\sf q}^a,{\sf q}^c)+\frac{1}{N} \delta^{ab} \delta^{cd} {\cal F}^{(2)}_g({\sf q}^a,{\sf q}^c)\Big] \Lambda_{\mu\nu}(P) \, ,
\ee
where the square bracket also reduces to $-{\cal P}^{ab,cd}$ in this limit of $\cQ\rightarrow 0$.

Adding up Eqs.~(\ref{piflofer}) and (\ref{piflobo}), the final result for the symmetric gluon self-energy at LO can be shown as
\ba\label{piflofinal}
\Pi^{ab,cd}_{F;\mu \nu}\big|_{{\rm LO}}(P,\cQ)&=&  i \frac{T}{p}  \Big\{ m_f^2\Big[\frac{1}{2}  \delta^{ad} \delta^{bc} \cot(\pi {\sf q}^{ac}) {\cal G}_f({\sf q}^a,{\sf q}^c)+\frac{1}{N} \delta^{ab} \delta^{cd} {\cal F}^{(2)}_f({\sf q}^a,{\sf q}^c)\Big]\,\nonumber\\
& + &m_g^2\Big[- \frac{1}{2}  \delta^{ad} \delta^{bc} \cot(\pi {\sf q}^{ac}) {\cal G}_g({\sf q}^a,{\sf q}^c)+\frac{1}{N} \delta^{ab} \delta^{cd} {\cal F}^{(2)}_g({\sf q}^a,{\sf q}^c)\Big]\Big\} \Lambda_{\mu\nu}(P) \, ,\nonumber \\
\ea
which is proportional to $g^2T^3/p$ and reproduces the correct behavior at vanishing background field,
\be\la{pif0}
\Pi^{ab,cd}_{F;\mu \nu}(P,\cQ\rightarrow 0)= - i \frac{T m_D^2}{p} \Lambda_{\mu\nu}(P){\cal P}^{ab,cd}\, .
\ee

Compared with the NLO result for the regarded gluon self-energy, both Eq.~(\ref{pirtotnlo}) and Eq.~(\ref{piflofinal}) are formally analogous to their counterparts at $\cQ=0$. The influence of the nonzero background field amounts to a modification on the Debye mass. However, such a modification turns out to be more complicated for the symmetric gluon self-energy where both trigonometric functions and the Bernoulli polynomials are included.

Given the fact that the LO symmetric gluon self-energy has a nonzero contribution at $\cQ=0$, we may consider it as a normal contribution. However, there exists an ambiguity because one can artificially introduce a term which vanishes at $\cQ=0$, and thus gives an anomalous contribution. Subtracting such a contribution from the original result, we can define the so-called normal contributions. It turns out to be interesting to make such a reorganization for the LO symmetric gluon self-energy. Detailed discussions can be found in Appendix~\ref{repif}.

\subsection{The next-to-leading order contributions in the HTL approximation}\label{secpifnlo}

It is also important to explore the NLO contributions to the symmetric gluon self-energy. Just like the LO $\Pi^{ab,cd}_{R;\mu \nu}(P,\cQ)$, these contributions are completely new as they only survive in a nonzero background field. More importantly, as we will show later, they are necessary to demonstrate the KMS condition at NLO in the HTL approximation.

There are three kinds of contributions arising at NLO. Dropping the term $\sim K_\mu K_\nu$ in Eq.~(\ref{pif}), the first kind of contribution can be obtained as
\ba\la{pifk1}
\Pi^{ab,cd}_{F;ij}(P,\cQ)\big|_{\rm I} &=& i \pi g^2 N_f \sum_{ef}{\cal P}^{ab,fe}{\cal P}^{cd,ef} \int \frac{d^3 {\bf k}}{(2\pi)^3} \frac{-{\hat p}_i {\hat k}_j-{\hat k}_i {\hat p}_j}{k}\Big[ \big(2 {\tilde n}_+^f(k) {\tilde n}_+^e(k)- {\tilde n}_+^f(k)\,\nonumber \\
&-& {\tilde n}_+^e(k)\big)\delta({\hat {\bf k}}\cdot {\hat {\bf p}}-p_0/p)+ \big(2 {\tilde n}_-^f(k) {\tilde n}_-^e(k)-{\tilde n}_-^f(k)- {\tilde n}_-^e(k)\big)\delta({\hat {\bf k}}\cdot {\hat {\bf p}}+p_0/p)\Big]\, .\nonumber \\
\ea
Here, we also take the spatial components as an example and other components in the Lorentz space can be discussed in a similar way. Notice that the term $\sim g_{\mu\nu} Q\cdot K$ does not appear in the above equation because it has no contribution due to the delta functions. In addition, the distribution function $ {\tilde n}(q_0,\cQ^f)$ is approximated to $ {\tilde n}(k_0,\cQ^f)$, and the subleading term in $\delta (Q^2)$ is neglected, namely $\delta (Q^2)\approx \delta({\hat {\bf k}}\cdot {\hat {\bf p}}\mp p_0/p)/(2kp)$ where the minus sign corresponds to the positive energy $k_0=k$ while the plus sign is for $k_0=-k$.

Performing the integral over the solid angle, we arrive at
\ba\la{pifk1new}
\Pi^{ab,cd}_{F;ij}(P,\cQ)\big|_{\rm I} &=& -i \frac{g^2}{2\pi}\frac{p_0}{p} N_f \frac{p_ip_j}{p^2}\theta(p^2-p_0^2)\sum_{ef} {\cal P}^{ab,fe}{\cal P}^{cd,ef}\int k dk \Big[ \big(2 {\tilde n}_+^f(k) {\tilde n}_+^e(k)- {\tilde n}_+^f(k)\,\nonumber \\
&-& {\tilde n}_+^e(k)\big)-\big(2 {\tilde n}_-^f(k) {\tilde n}_-^e(k)-{\tilde n}_-^f(k)- {\tilde n}_-^e(k)\big)\Big]\, .
\ea
We should mention the above result is not proportional to $\Lambda_{ij}(P)$ although one may naively expect such a proportionality.

The second kind of contribution at NLO comes from the following expansion of the distribution function for soft $p_0$,
\be
{\tilde n}(\pm k-p_0,\cQ^f)= {\tilde n}_{\pm}^f(k)\mp {\tilde n}_{\pm}^f(k)( {\tilde n}_{\pm}^f(k)-1)\frac{p_0}{T}+\cdots \, .
\ee
Picking up the subleading term which is suppressed by a factor of $p_0/T$, the resulting NLO contribution reads
\ba\la{pifk2}
\Pi^{ab,cd}_{F;ij}(P,\cQ)\big|_{\rm II} &=&- i 2 \pi g^2 N_f \sum_{ef} {\cal P}^{ab,fe}{\cal P}^{cd,ef}\frac{p_0}{T}\int \frac{d^3 {\bf k}}{(2\pi)^3} \frac{{\hat k}_i {\hat k}_j}{p}\Big\{ \big[2 {\tilde n}_+^f(k)( {\tilde n}_+^f(k)-1) {\tilde n}_+^e(k)\,\nonumber \\
&-& {\tilde n}_+^f(k)( {\tilde n}_+^f(k)-1)\big]\delta({\hat {\bf k}}\cdot {\hat {\bf p}}-p_0/p)- \big[2 {\tilde n}_-^f(k)({\tilde n}_-^f(k)-1) {\tilde n}_-^e(k)\,\nonumber \\
&-&{\tilde n}_-^f(k)({\tilde n}_-^f(k)-1)\big]\delta({\hat {\bf k}}\cdot {\hat {\bf p}}+p_0/p)\Big\}\, ,
\ea
where we keep $\sim K_\mu K_\nu$ in Eq.~(\ref{pif}) and other terms such as $\sim K_\mu P_\nu$ and $\sim K_\nu P_\mu$ should be dropped which lead to higher order contributions. For the same reason, we also take $\delta (Q^2)\approx \delta({\hat {\bf k}}\cdot {\hat {\bf p}}\mp p_0/p)/(2kp)$ in the above equation.

The integral over the solid angle is the same as the LO contribution, therefore, the second kind of contribution at NLO is proportional to $\Lambda_{ij}(P)$. Then, we can show that
\ba\la{pifk2new}
\Pi^{ab,cd}_{F;ij}(P,\cQ)\big|_{\rm II} &=& i \frac{g^2}{4\pi^2}\frac{p_0}{p} N_f\Lambda_{ij}(P)\sum_{ef} {\cal P}^{ab,fe}{\cal P}^{cd,ef}\int k^2 dk\Big(\frac{\partial  {\tilde n}_+^f(k)}{\partial k}-2 \frac{\partial  {\tilde n}_+^f(k)}{\partial k} {\tilde n}_+^e(k)\,\nonumber \\
&-&\frac{\partial  {\tilde n}_-^f(k)}{\partial k}+2 \frac{\partial  {\tilde n}_-^f(k)}{\partial k} {\tilde n}_-^e(k)\Big)\, .
\ea

Now we turn to the third kind of contribution which takes a very similar form as Eq.~(\ref{pifij}),
\ba\la{pifk3}
\Pi^{ab,cd}_{F;ij}(P,\cQ)\big|_{\rm III}  &=& i 2 \pi g^2 N_f \sum_{ef} {\cal P}^{ab,fe}{\cal P}^{cd,ef}\int \frac{d^3 {\bf k}}{(2\pi)^3} \frac{ {\hat k}_i  {\hat k}_j}{ p}\Big\{ [2 {\tilde n}_+^f(k) {\tilde n}_+^e(k)- {\tilde n}_+^f(k)- {\tilde n}_+^e(k)]\,\nonumber \\
&\times&\delta^{{\rm NLO}}\Big({\hat {\bf k}}\cdot {\hat {\bf p}}-p_0/p+\frac{P^2}{2kp}\Big)+ [2 {\tilde n}_-^f(k) {\tilde n}_-^e(k)-{\tilde n}_-^f(k)- {\tilde n}_-^e(k)]\,\nonumber\\
&\times&\delta^{{\rm NLO}}\Big({\hat {\bf k}}\cdot {\hat {\bf p}}+p_0/p+\frac{P^2}{2kp}\Big)\Big\}\, .
\ea
As we can see the only difference is that in Eq.~(\ref{pifk3}), we keep the subleading term $P^2/(2kp)$ in the delta functions where the superscript ``NLO" means that only the NLO contributions should be kept in the integral over the solid angle, namely,
\ba
\int d \Omega {\hat k}_i  {\hat k}_j \delta \Big({\hat {\bf k}}\cdot {\hat {\bf p}}\pm p_0/p+\frac{P^2}{2kp}\Big) &\equiv& \int d \Omega {\hat k}_i  {\hat k}_j \delta^{{\rm LO}}\Big({\hat {\bf k}}\cdot {\hat {\bf p}}\pm p_0/p+\frac{P^2}{2kp}\Big)\,\nonumber \\
& +&\int d \Omega {\hat k}_i  {\hat k}_j \delta^{{\rm NLO}}\Big({\hat {\bf k}}\cdot {\hat {\bf p}}\pm p_0/p+\frac{P^2}{2kp}\Big)\,,
\ea
where the first integral on the right-hand side of the above equation is actually given by Eq.~(\ref{inso}) where the term $P^2/(2 k p)$ is not relevant, while the second one is given by
\be\la{inso2}
\int d \Omega {\hat k}_i  {\hat k}_j \delta^{{\rm NLO}}\Big({\hat {\bf k}}\cdot {\hat {\bf p}}\pm p_0/p+\frac{P^2}{2kp}\Big) =\pm\frac{p_0}{k}\Big( \Lambda_{ij}(P) - 2  \pi \frac{p_i p_j}{p^2}\theta(p^2-p_0^2)\Big)\, .
\ee
Using Eq.~(\ref{inso2}), we can express $\Pi^{ab,cd}_{F;ij}(P,\cQ)\big|_{\rm III}$ as
\ba\la{pifk3new}
&&\Pi^{ab,cd}_{F;ij}(P,\cQ)\big|_{\rm III}  = -i \frac{g^2}{4\pi^2}\frac{p_0}{p} N_f \Big(\Lambda_{ij}(P)-2\pi\frac{p_ip_j}{p^2}\theta(p^2-p_0^2)\Big)\sum_{ef} {\cal P}^{ab,fe}{\cal P}^{cd,ef}\int k dk \,\nonumber \\
&&\quad\times\Big[ \big(2 {\tilde n}_+^f(k) {\tilde n}_+^e(k)- {\tilde n}_+^f(k)- {\tilde n}_+^e(k)\big)-\big(2 {\tilde n}_-^f(k) {\tilde n}_-^e(k)-{\tilde n}_-^f(k)- {\tilde n}_-^e(k)\big)\Big]\, .
\ea
Interestingly, we find that the first kind of contribution as given in Eq.~(\ref{pifk1new}) is completely canceled by the above equation and the remaining terms are simply proportional to $\Lambda_{ij}(P)$. By combining Eqs.~(\ref{pifk1new}), (\ref{pifk2new}), and (\ref{pifk3new}), the symmetric gluon self-energy at NLO takes the following form:
\ba\la{pifnlo}
\Pi^{ab,cd}_{F;\mu\nu}(P,\cQ)\big|_{\rm NLO}  &=& - i \frac{g^2}{4\pi^2}\frac{p_0}{p} N_f\Lambda_{\mu \nu}(P)\sum_{ef} {\cal P}^{ab,fe}{\cal P}^{cd,ef}\int k^2 dk\bigg[\Big(\frac{\partial  {\tilde n}_+^f(k)}{\partial k} {\tilde {\cal N}}_+^e(k)\nonumber \\
&-& \frac{\partial  {\tilde n}_+^e(k)}{\partial k} {\tilde {\cal N}}_+^f(k)\Big)-\Big(\frac{\partial  {\tilde n}_-^f(k)}{\partial k} {\tilde {\cal N}}_-^e(k)- \frac{\partial  {\tilde n}_-^e(k)}{\partial k} {\tilde {\cal N}}_-^f(k)\Big)\bigg]\, ,
\ea
where ${\tilde {\cal N}}_\pm^f(k)={\tilde n}_\pm^f(k)-1/2$. The other Lorentz components can be obtained in a similar way, therefore, the corresponding details are omitted here. Notice that for the ``00 " component, the first two kinds of contributions are both proportional to $\Lambda_{00}(P)$, while the third kind of contribution is not involved in NLO calculation.

According to Eq.~(\ref{pifnlo}), it is clear to see that the NLO contribution in $\Pi^{ab,cd}_{F;\mu\nu}(P,\cQ)$ vanishes when $\cQ=0$ or $\cQ^e=\cQ^f$. This is analogous to the retarded gluon self-energy at LO. Therefore, no diagonal component exists and $\Pi^{ab,cd}_{F;\mu\nu}(P,\cQ)\big|_{\rm NLO}$ is considered as an anomalous contribution.

We can further integrate over $k$ in Eq.~(\ref{pifnlo}) by making use of the following equation:
\ba
&&\frac{1}{T^2}\int k^2 dk\Big(\frac{\partial  {\tilde n}_+^f(k)}{\partial k} {\tilde {\cal N}}_+^e(k)- \frac{\partial  {\tilde n}_+^e(k)}{\partial k} {\tilde {\cal N}}_+^f(k)\Big) = i \cot(\pi {\sf q}^{fe}) \big[{\rm Li}_2(- e^{i 2 \pi {\sf q}^{f}})+{\rm Li}_2(- e^{i 2 \pi {\sf q}^{e}})\big]\, \nonumber \\
&&\quad- \csc^2 (\pi {\sf q}^{fe}) \big[{\rm Li}_3(- e^{i 2 \pi {\sf q}^{f}})-{\rm Li}_3(- e^{i 2 \pi {\sf q}^{e}})\big]\, ,
\ea
where the polylogarithm functions can be related to the Bernoulli polynomials via
\ba
{\rm Li}_2(- e^{i 2 \pi {\sf q}^{f}})+{\rm Li}_2(- e^{- i 2 \pi {\sf q}^{f}}) &=& 2\pi^2 B_2({\tilde{\sf q}}^f)\, ,\nonumber \\
{\rm Li}_3(- e^{i 2 \pi {\sf q}^{f}})-{\rm  Li}_3(- e^{- i 2 \pi {\sf q}^{f}}) &=& i\frac{4\pi^3}{3} B_3({\tilde{\sf q}}^f)\, .
\ea
Consequently, the final result for the $\Pi^{ab,cd}_{F;\mu\nu}(P,\cQ)$ at NLO in the HTL approximation can be expressed as
\ba\la{pifnlofinal}
&&\Pi^{ab,cd}_{F;\mu\nu}(P,\cQ)\big|_{\rm NLO}
=  \frac{p_0}{2p}\delta^{ad}\delta^{bc}\Big[m_f^2 \Big(\frac{1}{2}\csc^2 (\pi {\sf q}^{ac}) {\cal G}_f( {\sf q}^a, {\sf q}^c)+\cot(\pi {\sf q}^{ac}){\cal F}_f^{(1)}( {\sf q}^a, {\sf q}^c)\Big)\nonumber \\
&&\quad+\,m_g^2\Big(-\frac{1}{2}\csc^2 (\pi {\sf q}^{ac}) {\cal G}_g( {\sf q}^a, {\sf q}^c)+\cot(\pi {\sf q}^{ac}){\cal F}_g^{(1)}( {\sf q}^a, {\sf q}^c)\Big)\Big]  \Lambda_{\mu \nu}(P)\, .
\ea

Given the above discussions about the quark loop diagram, calculations for the pure gauge part are straightforward, although rather tedious. We only provide the corresponding result which is given in the second line of the above equation. As compared to the LO result in Eq.~(\ref{piflofinal}) which is pure imaginary, the NLO contribution is real and suppressed by a factor of $p_0/T$. We would like to mention that although the NLO symmetric gluon self-energy is also an anomalous contribution as it vanishes at $\cQ=0$, unlike the LO retarded gluon self-energy, it is transverse due to the vanishment of $P^\mu  \Lambda_{\mu \nu}(P)$. On the other hand, a similar identity as Eq.~(\ref{idsum}) also holds for the NLO symmetric gluon self-energy.

Finally, a vanishing result from Eq.~(\ref{pifnlofinal}) is required in the special case where $\cQ=0$ or $\cQ^a=\cQ^c$. However, this is not very obvious based on the above given expressions. In fact, taking ${\sf q}^a\rightarrow {\sf q}^c \pm \epsilon$ and making a Taylor expansion for infinitely small $\epsilon$, both of the two terms associated with different trigonometric functions in the square bracket have contributions at successive orders $\sim 1/\epsilon$, $\sim \epsilon^{0}$, $\sim \epsilon$ and so on. However, the divergent and finite parts in the Taylor series are exactly canceled, leaving a vanishing result as expected.

\section{The Kubo-Martin-Schwinger condition in a semi-quark-gluon plasma}
\la{kms}

With the HTL gluon self-energies computed up to NLO in Secs.~\ref{pire} and \ref{pisy}, we can study the KMS condition in a semi-quark-gluon plasma. As a natural extension of the KMS condition at $\cQ=0$, it can be expressed as
\be\label{kmsbf}
\Pi^{ab,cd}_{F;\mu \nu}(P,\cQ) = \big(1+2 n (p_0, \cQ^{ab})\big) {\rm sgn} (p_0) \big(\Pi^{ab,cd}_{R;\mu \nu}(P,\cQ)-\Pi^{ab,cd}_{A;\mu \nu}(P,\cQ) \big)\, .
\ee
A formal derivation of the above equation can be carried out by following a similar procedure as used in the ${\cal Q}=0$ case. As already mentioned before, the introduced background field acts like an imaginary chemical potential. In analogy to an ordinary chemical potential, it only brings a ${\cal Q}$ dependence in the distribution function as given in Eqs.~(\ref{bd}) and (\ref{fd}).

For timelike gluons, studying the KMS condition is rather trivial because the symmetric gluon self-energy vanishes due to the $\theta$ function and meanwhile the retarded and advanced ones become identical. In the following, we consider only the spacelike gluons with $p_0^2<p^2$.

Let us first recall what happens in zero background field. To verify Eq.~(\ref{kmsbf}), one can make use of the following relation among the dimensionless functions, \footnote{$\Gamma^{(2)}_{R, \mu\nu}(P)$ is given by Eq.~(\ref{defg2}), while $\Gamma^{(2)}_{A, \mu\nu}(P)$ can be obtained by changing $+i \epsilon$ into $-i \epsilon$ in the same equation.}
\be\la{redf}
\Gamma^{(2)}_{R, \mu\nu}(P)-\Gamma^{(2)}_{A, \mu\nu}(P)= i \frac{p_0} {2p} \Lambda_{\mu\nu}(P) \, ,
\ee
which can be obtained via the identity
\be
\ln\frac{p_0+p+i \epsilon}{p_0-p+i \epsilon}-\ln\frac{p_0+p - i \epsilon}{p_0-p -i \epsilon} =-2\pi i \,\quad{\rm for}\quad p_0^2<p^2\, .
\ee

Furthermore, the external momentum $p_0$ is considered to be soft  in the HTL approximation, {\em i.e.}, $p_0\ll T$, and the bosonic distribution function can be expanded as
\be\la{exdis}
(1+2 n (p_0, \cQ= 0)) {\rm sgn} (p_0) \approx \frac{2T}{p_0}+\frac{p_0}{6T}+\cdots \, ,
\ee
where higher order terms are power suppressed for $p_0/T \ll 1$ and should be dropped for consistency if only the LO (nonvanishing) contributions in the gluon self-energies are taken into account. Then, it is easy to show the validity of Eq.~(\ref{kmsbf}) based on Eqs.~(\ref{pir0}) and (\ref{pif0}). Notice that the bosonic distribution function is enhanced by $\sim T/p_0$, therefore, a $T^3$ behavior can be found on the right-hand side of Eq.~(\ref{kmsbf}) which matches the $T$ dependence in the symmetric gluon self-energy.

It is certainly interesting to ask if the KMS condition can hold when $\cQ \neq 0$. For the diagonal components of the gluon self-energies, Eq.~(\ref{kmsbf}) becomes
\be\label{kmsdia}
\Pi^{aa,cc}_{F;\mu \nu}(P,\cQ) = (1+2 n (p_0, \cQ^{aa}=0)) {\rm sgn} (p_0) (\Pi^{aa,cc}_{R;\mu \nu}(P,\cQ)-\Pi^{aa,cc}_{A;\mu \nu}(P,\cQ) )\, .
\ee
Given the explicit results of the HTL gluon self-energies, the above equation is valid because the diagonal components of the gluon self-energies at $\cQ\neq 0$ can be simply obtained from those at $\cQ=0$ by multiplying a $\cQ$-dependent function which is the same for the retarded/advanced and symmetric solutions, see Eqs.~(\ref{pirtotnlo}) and (\ref{piflofinal}). In addition, the distribution function for diagonal gluons is not affected by the background field, therefore, Eq.~(\ref{exdis}) can apply. It is worthwhile to mention that for the symmetric gluon self-energy, the diagonal components are taken from the LO contribution, while for the retarded/advanced one, they arise at NLO in the HTL approximation. This is actually in analogy with that in zero background field.

For off-diagonal components, the KMS condition reads
\be\label{kmsoff}
\Pi^{ac,ca}_{F;\mu \nu}(P,\cQ) = (1+2 n (p_0, \cQ^{ac})) {\rm sgn} (p_0) (\Pi^{ac,ca}_{R;\mu \nu}(P,\cQ)-\Pi^{ac,ca}_{A;\mu \nu}(P,\cQ) )\, ,
\ee
where the bosonic distribution function becomes $\cQ$ dependent. As a result, there is no enhancement in $n (p_0, \cQ^{ac})$ when $p_0$ is small. Instead, for $p_0/T \ll 1$ we find that
\be\label{offdis}
\big(1+2 n (p_0, \cQ^{ac})\big) {\rm sgn} (p_0) = i \cot (\pi {\sf q}^{ac})+ \frac{p_0}{2T} \csc^2 (\pi {\sf q}^{ac}) + \cdots \, , \quad {\sf q}^a\neq {\sf q}^c\, .
\ee
For later use, we write
\ba\la{offdis1}
(1+2 n (p_0, \cQ^{ac})) {\rm sgn} (p_0) \big|_{{\rm LO}} &\equiv&  i \cot (\pi {\sf q}^{ac}) \, , \quad {\sf q}^a\neq {\sf q}^c\, , \\
(1+2 n (p_0, \cQ^{ac})) {\rm sgn} (p_0) \big|_{{\rm NLO}} &\equiv& \frac{p_0}{2T} \csc^2 (\pi {\sf q}^{ac}) \, ,\quad {\sf q}^a\neq {\sf q}^c\, .
\ea
Given the above expansion, it is not a surprise to see the trigonometric functions in our result for $\Pi^{ab,cd}_{F;\mu\nu}(P,\cQ)$ as they also show up in the KMS condition.

The off-diagonal components of $\Pi^{ab,cd}_{F;\mu \nu}(P,\cQ)$ at LO are proportional to $T^3$ as shown in Eq.~(\ref{piflofinal}). Clearly, to show the KMS condition, one needs to include only the LO terms in the retarded/advanced gluon self-energies in order to match the power of $T$. Using Eq.~(\ref{pirtotlo}), we get
\ba\label{ralo}
\Pi^{ac,ca}_{R;\mu \nu}  \big|_{{\rm LO}}(P,\cQ)&-&\Pi^{ac,ca}_{A;\mu \nu}  \big|_{{\rm LO}}(P,\cQ) \nonumber \\
&=& i \frac{T}{p_0} \big[m_f^2 {\cal G}_f({\sf q}^a,{\sf q}^c)  - m_g^2 {\cal G}_g({\sf q}^a,{\sf q}^c) \big] \big[ \Gamma^{(1)}_{R, \mu\nu}(P)- \Gamma^{(1)}_{A, \mu\nu}(P)\big]\, \nonumber \\
&=&  \frac{T}{2 p} \big[m_f^2 {\cal G}_f({\sf q}^a,{\sf q}^c) - m_g^2 {\cal G}_g({\sf q}^a,{\sf q}^c) \big] \Lambda_{\mu\nu}(P)   \, , \quad {\sf q}^a\neq {\sf q}^c\, .
\ea
Taking also the LO term in the expansion of the $\cQ$-dependent distribution function, according to Eq.~(\ref{offdis}), it can be shown that
\ba
&&(1+2 n (p_0, \cQ^{ac})) {\rm sgn} (p_0) (\Pi^{ac,ca}_{R;\mu \nu}  (P,\cQ)-\Pi^{ac,ca}_{A;\mu \nu} (P,\cQ))\nonumber \\ & \xlongequal{{\rm LO}}& (1+2 n (p_0, \cQ^{ac})) {\rm sgn} (p_0) \big|_{{\rm LO}} (\Pi^{ac,ca}_{R;\mu \nu}  \big|_{{\rm LO}}(P,\cQ)-\Pi^{ac,ca}_{A;\mu \nu}  \big|_{{\rm LO}}(P,\cQ))\nonumber \\
& =&  i \frac{T}{2p}\cot (\pi {\sf q}^{ac}) \big[m_f^2 {\cal G}_f({\sf q}^a,{\sf q}^c) - m_g^2 {\cal G}_g({\sf q}^a,{\sf q}^c) \big]  \Lambda_{\mu\nu}(P)   \,, \quad {\sf q}^a\neq {\sf q}^c\, ,
\ea
which is nothing but the off-diagonal components of $\Pi^{ab,cd}_{F;\mu \nu}  \big|_{{\rm LO}}(P,\cQ)$ as given in Eq.~(\ref{piflofinal}). Thus, the KMS condition in Eq.~(\ref{kmsoff}) is explicitly verified at LO. Here, the involved contributions in both symmetric and retarded/advanced gluon self-energies come from the same order in the HTL approximation. This is very different from the KMS condition for diagonal components where the LO terms in $\Pi^{aa,cc}_{R/A;\mu \nu}(P,\cQ)$ vanish and the corresponding NLO contributions $\sim T^2$ can be related to the LO $\Pi^{aa,cc}_{F;\mu \nu} (P,\cQ)$ via Eq.~(\ref{kmsdia}) thanks to the enhanced bosonic distribution function.

According to the above discussions, when considering only the LO contributions in the symmetric gluon self-energy, the verification of KMS conditions requires a computation of $\Pi^{ab,cd}_{R/A;\mu \nu} (P,\cQ)$ up to NLO in the HTL approximation. However, the off-diagonal components in the NLO $\Pi^{ab,cd}_{R/A;\mu \nu} (P,\cQ)$ which are related to the LO symmetric gluon self-energy via the KMS condition when $\cQ=0$ are missing in our discussion for nonzero background field. In fact, with $\cQ \neq 0$, these missing terms satisfy the following equation:
\ba
 \Pi^{ac,ca}_{R;\mu \nu}  \big|_{{\rm NLO}}(P,\cQ)&-&\Pi^{ac,ca}_{A;\mu \nu}  \big|_{{\rm NLO}}(P,\cQ)\nonumber \\
 &=&   -\big[ m_f^2 {\cal F}^{(1)}_f({\sf q}^a,{\sf q}^c)+m_g^2 {\cal F}^{(1)}_g({\sf q}^a,{\sf q}^c)\big]\big[\Gamma^{(2)}_{R,\mu\nu}(P)-\Gamma^{(2)}_{A,\mu\nu}(P)\big]\,\nonumber\\
&=&- i \frac{p_0}{2p} \big[ m_f^2 {\cal F}^{(1)}_f({\sf q}^a,{\sf q}^c)+m_g^2 {\cal F}^{(1)}_g({\sf q}^a,{\sf q}^c)\big] \Lambda_{\mu\nu}(P)   \,, \,\, {\sf q}^a\neq {\sf q}^c\, .
\ea
As compared to Eq.~(\ref{ralo}), the above contribution is suppressed by a factor of  $p_0/T$ and turns out to be useful to verify the KMS condition Eq.~(\ref{kmsoff}) at NLO.

As suggested by Eq.~(\ref{offdis}), there are two sources that contribute when studying the KMS condition Eq.~(\ref{kmsoff}) at NLO. Explicitly, we can show that
\ba\label{kmsrnlo}
&&(1+2 n (p_0, \cQ^{ac})) {\rm sgn} (p_0) (\Pi^{ac,ca}_{R;\mu \nu}  (P,\cQ)-\Pi^{ac,ca}_{A;\mu \nu} (P,\cQ))\nonumber \\ & \xlongequal{{\rm NLO}}& (1+2 n (p_0, \cQ^{ac})) {\rm sgn} (p_0) \big|_{{\rm LO}} (\Pi^{ac,ca}_{R;\mu \nu}  \big|_{{\rm NLO}}(P,\cQ)-\Pi^{ac,ca}_{A;\mu \nu}  \big|_{{\rm NLO}}(P,\cQ))\nonumber \\
&+& (1+2 n (p_0, \cQ^{ac})) {\rm sgn} (p_0) \big|_{{\rm NLO}} (\Pi^{ac,ca}_{R;\mu \nu}  \big|_{{\rm LO}}(P,\cQ)-\Pi^{ac,ca}_{A;\mu \nu}  \big|_{{\rm LO}}(P,\cQ))\nonumber \\
& =&   \frac{p_0}{2p}\cot (\pi {\sf q}^{ac}) \big[ m_f^2 {\cal F}^{(1)}_f({\sf q}^a,{\sf q}^c)+m_g^2 {\cal F}^{(1)}_g({\sf q}^a,{\sf q}^c)\big] \Lambda_{\mu\nu}(P) \nonumber \\
& +&   \frac{p_0}{4p} \csc^2 (\pi {\sf q}^{ac}) \big[m_f^2 {\cal G}_f({\sf q}^a,{\sf q}^c)  - m_g^2 {\cal G}_g({\sf q}^a,{\sf q}^c) \big] \Lambda_{\mu\nu}(P)    \,, \quad {\sf q}^a\neq {\sf q}^c\, .
\ea
By comparing Eq.~(\ref{kmsrnlo}) with Eq.~(\ref{pifnlofinal}) which is obtained by a direct calculation of $\Pi^{ab,cd}_{F;\mu \nu}  \big|_{{\rm NLO}}(P,\cQ)$, the validity of Eq.~(\ref{kmsoff}) at NLO in the HTL approximation is verified.

When we expand the distribution function in Eq.~(\ref{offdis}) with $p_0\ll T$, we implicitly assume $\cQ \sim T$, so that the dimensionless background field ${\sf q} \sim {\cal O}(1)$. Notice that a special case where ${\sf q} \sim p_0/T$ leads to a different expansion of the distribution function which can be expressed as
\be\la{dissc}
\big(1+2 n (p_0, \cQ^{ac})\big) {\rm sgn} (p_0) = \frac{1}{1-i\,2 \pi {\hat {\sf q}}^{ac}} \frac{2T}{p_0} + \cdots \, , \quad {\rm for}\quad{\sf q}^a\neq {\sf q}^c\, ,
\ee
where ${\sf q}^a/{\hat {\sf q}}^a = p_0/T$ with ${\hat {\sf q}}^a \sim {\cal O}(1)$ and higher order terms suppressed by powers of $p_0/T$ are dropped. To verify the KMS condition for the off-diagonal components in this special case, we need to also expand the Bernoulli polynomials in the gluon self-energies for small $p_0/T$. According to Eq.~(\ref{pirtotlo}), the LO contribution to the retarded solution becomes
\be
\Pi^{ac,ca}_{R;\mu \nu}  \big|_{{\rm LO}}(P,\cQ)=- i\, 2\pi {\hat {\sf q}}^{ac}  m_D^2 \Gamma^{(1)}_{\mu\nu}(P)+\cdots\, ,\quad{\rm for}\quad {\sf q} \sim p_0/T\ll 1\,.
\ee
At the same order $\sim g^2 T^2$, there is another contribution which can be obtained from Eq.~(\ref{pirtotnlo}) by setting $\cQ=0$. Therefore, the final result reads
\be
\Pi^{ac,ca}_{R;\mu \nu} (P,\cQ)=-  m_D^2 \big(i\, 2\pi  {\hat {\sf q}}^{ac} \Gamma^{(1)}_{\mu\nu}(P)+ \Gamma^{(2)}_{\mu\nu}(P)\big)+\cdots\, ,\quad{\rm for}\quad {\sf q} \sim p_0/T\ll 1\,.
\ee
Based on the above equation, it is easy to show
\be\la{sc}
\Pi^{ac,ca}_{R;\mu \nu} (P,\cQ)-\Pi^{ac,ca}_{A;\mu \nu} (P,\cQ)=-i \frac{p_0}{2p} m_D^2 (1- i\, 2\pi {\hat {\sf q}}^{ac}) \Lambda_{\mu\nu}(P)\, .
\ee
On the other hand, the symmetric gluon self-energy is simply given by Eq.~(\ref{pif0}) when expanding Eq.~(\ref{piflofinal}) for ${\sf q} \sim p_0/T\ll 1$ and dropping the higher order contributions. Together with Eqs.~(\ref{dissc}) and (\ref{sc}), the KMS condition is proved to hold in the special case where the background field is at order $\sim p_0/T$.

It is interesting to point out that the anomalous contribution in the retarded/advanced gluon self-energy is $\sim g^2T^3/p_0$ when the background field $\cQ \sim T$. However, for small $\cQ$ where $\cQ \sim p_0$ or ${\sf q} \sim p_0/T$, the anomalous contribution becomes comparable to $\Pi_{R/A}$ at $\cQ=0$ which is proportional to the Debye mass square. As a result, the $\cQ$ modification on the retarded/advanced solution is not negligible even for small background field. On the contrary, for the symmetric gluon self-energy, such a $\cQ$ modification is power suppressed as compared to the result at vanishing background field.

To conclude, our discussion confirms that in general, using the real-time gluon self-energies computed in the HTL approximation, the KMS condition is satisfied order by order for $p_0/T\ll 1$.  As demonstrated in this work, it is a highly nontrivial extension from $\cQ=0$ to $\cQ \neq 0$ due to some new features arising in the off-diagonal components.

\section{Conclusions and Outlook}\la{conclusions}

We computed the one-loop gluon self-energy up to the next-to-leading order in the HTL approximation where a background field $\cQ$ has been introduced for the vector potential, leading to a nontrivial expectation value for the Polyakov loop in the deconfined phase. The explicit results for the retarded/advanced and symmetric gluon self-energies in the Keldysh representation have been obtained.

Some new terms which only survived at nonvanishing background ground were found in our computations. For the retarded/advanced gluon self-energy, they came from the LO contributions which were proportional to $g^2 T^3/p_0$, enhanced by a factor of $T/p_0$ as compared to the results at $\cQ =0$. For the symmetric gluon self-energy, on the other hand, these new terms arose at NLO and were suppressed by $p_0/T$ as compared to the LO contributions. It is worth pointing out that these anomalous contributions are proportional to $\delta^{ad}\delta^{bc}$ and antisymmetric when flipping the color indices $a\leftrightarrow c$, therefore, Eq.~(\ref{idsum}) is valid simply because the projection operator ${\cal P}^{ac,ca}$ is symmetric under $a\leftrightarrow c$. In addition, the NLO contributions in $\Pi_{R/A}$ as well as the LO contribution in $\Pi_F$ are formally analogous to the well-known result computed in the completely deconfined QGP. The influence of the nonzero background field merely amounts to a modification on the Debye mass. According to Eqs.~(\ref{pirtotnlo}) and (\ref{piflofinal}), we found that the fermionic modification differs from the bosonic one and such a modification is also different for the retarded/advanced and symmetric solutions of the gluon self-energies. In the limit $\cQ \rightarrow 0$, both of these contributions reproduce the desired results as shown in Eqs.~(\ref{pir0}) and (\ref{pif0}).

With the obtained results, we also verified that the KMS condition can be satisfied in a semi-QGP with nonzero background field. It is a nontrivial extension from $\cQ=0$ to $\cQ\neq 0$, especially for the off-diagonal components where the statistic distributions for the soft gluons depend on the background field, hence have no Bose enhancement $\sim T/p_0$. In this case, we found that the anomalous contributions played an important role which guaranteed the order-by-order satisfaction of the KMS condition in the HTL approximation.

In this work, all the computations were carried out within the HTL perturbation theory where however, nonzero values of $\cQ$ cannot be consistently generated because the corresponding equations of motion of the background field require the system to always stay in the completely deconfined phase where $\cQ$ vanishes. To quantitatively study the background field modification on the gluon self-energies, one needs to specify the values of $\cQ$ at a given temperature by using either the lattice simulations on the Polyakov loops, or the equations of motion based on effective theories in the semi-QGP. In fact, some other unphysical behaviors have also been found in recent works where only perturbative contributions were taken into account\cite{Guo:2020jvc,KorthalsAltes:2019yih,KorthalsAltes:2020ryu}. Naively, one may relate these problems to the nontransversality of the retarded gluon self-energy. A further analysis\cite{KorthalsAltes:2020ryu} has shown that gauge invariant sources, which are nonlinear in the gauge potential $A_0$, generate a novel constrained contribution to the gluon self-energy in the perturbation theory. It exactly cancels the nontransverse term $\sim M_{\mu}M_{\nu}$, and thus the total gluon self-energy remains transverse. It is obvious that the KMS condition is not affected by dropping this nontransverse term, however, for any gauge invariant source, it was found that there exists an unexpected discontinuity in the free energy appearing at order $\sim g^3$ as the background field vanishes. In addition, including the anomalous contribution, either with or without the nontransverse term, one encountered an ill-defined static limit in the resummed gluon propagator\cite{Guo:2020jvc}. Therefore, developing effective theories to study the physics in semi-QGP appears to be necessary.

In Ref.~\cite{Hidaka:2020vna}, a new contribution coming from two-dimensional ghost fields embedded isotropically into four dimensions was added to the perturbative gluon self-energy. As a result, the LO anomalous contributions in Eq.~(\ref{pirtotlo}) have been completely canceled, leading to an effective retarded gluon self-energy which is given by Eq.~(\ref{effpir}) with the $\cQ$-dependent mass squared shifted by a dimensional constant. 
If we naively assume the same also happens to the symmetric gluon self-energy, \footnote{In this case, all the anomalous contributions from Eqs.~(\ref{pifano}), (\ref{pifano2}) and (\ref{pifnlofinal}) should be dropped and the effective symmetric gluon self-energy is given by Eq.~(\ref{effpif}) where the same constant shift applies to the $\cQ$-dependent mass squared.} then it can be shown that the effective gluon self-energies satisfy the KMS condition provided that the background field only affects the statistical distributions of the hard particles, in other words, $(1+2 n (p_0, \cQ^{ab})) {\rm sgn} (p_0)$ should be approximated by $2T/p_0$ even for the off-diagonal components because the distribution function here is for the soft gluons. However, we have to admit that there seems no physical reason to make such an assumption. On the other hand, acknowledging the effective retarded gluon self-energy proposed in Ref.~\cite{Hidaka:2020vna}, it turns out to be necessary to ignore the $\cQ$ dependence in the distribution functions of the soft particles. Suppose that Eq.~(\ref{offdis1}) was used, the off-diagonal components of the resulting effective $\Pi^{ac,ca}_{F;\mu \nu} (P,\cQ)$ would become antisymmetric when we flip the color indices $a$ and $c$. This is obviously contradictory to the desired result when taking the limit $\cQ \rightarrow 0$.

The current work is an important step towards a full study on the gluon self-energy and other related physics in a semi-QGP. Properly incorporating the nonperturbative contributions based on effective theories, in particular, when the model in Ref.~\cite{Hidaka:2020vna} is adopted, how the two-dimensional ghost fields would affect the symmetric gluon self-energy is obviously an interesting topic which still needs to be further investigated.

\section*{Acknowledgements}
The work is supported by the NSFC of China under Projects No. 12065004 and No. 12147211.

\appendix

\section{CONTRIBUTIONS FROM PURE GAUGE PART TO THE REAL-TIME GLUON SELF-ENERGIES }\label{app}

In this Appendix, we provide some details about the calculations of the real-time gluon self-energies in a pure gauge theory. We adopt the double line notation and focus on the color structures of each Feynman diagram. The outcome will explicitly show that the total contribution from the pure gauge part is similar to the quark-loop diagram which is also in analogy with what happened in the zero background field.

\begin{figure}[hbtp]
\centering
\includegraphics[width=0.5\textwidth]{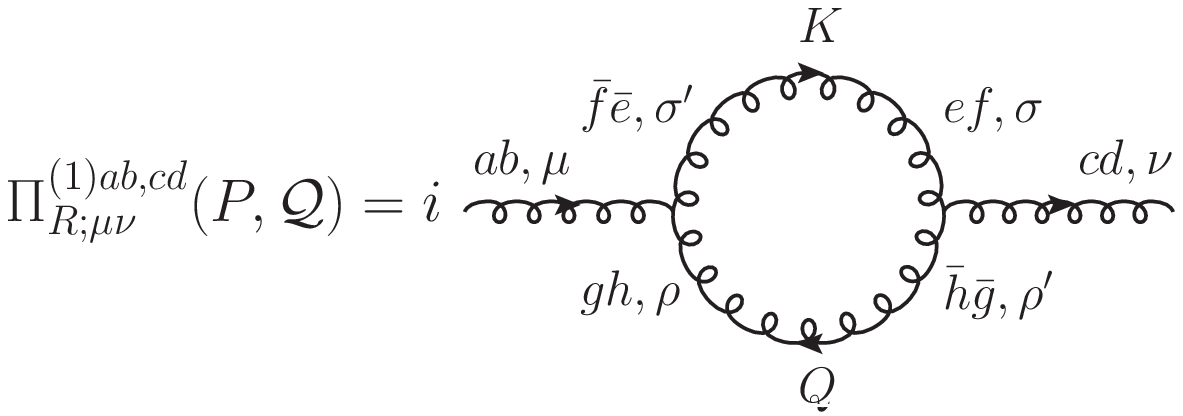}
\vspace*{-0.2cm}
\caption{\label{gloop}
Gluon-loop diagram contributing to the one-loop gluon self-energy.}
\end{figure}

We start by considering the gluon-loop diagram. Using the Feynman rules as given in Sec.~\ref{feynman}, it can be shown that\footnote{To be more clear about the Lorentz and color indices used in Eq.~(\ref{pirglue}), one can refer to Fig.~\ref{gloop}.}
\ba\la{pirglue}
\Pi^{(1) ab,cd}_{R;\mu \nu}(P,\cQ)&=&\Pi^{(1)ab,cd}_{11;\mu \nu}(P,\cQ)+\Pi^{(1)ab,cd}_{12;\mu \nu}(P,\cQ)\nonumber \\
&=& \frac{i}{2} \sum_{{\rm colors}}\int\frac{d^4 K}{(2\pi)^4} \big[D_{11}(Q,\cQ^{gh})D_{11}(K,\cQ^{ef})-D_{21}(Q,\cQ^{gh})D_{12}(K,\cQ^{ef})\big]\,\nonumber \\
&\times& (-g^{\sigma \sigma^\prime})(-g^{\rho \rho^\prime}){\cal V}_{{\rm L},\,\mu\rho\sigma}^{ab,{\bar e} {\bar f},gh}(P,Q,K){\cal V}_{{\rm R},\, \nu\rho^\prime\sigma^\prime}^{cd,{\bar g} {\bar h},ef}(P,Q,K) {\cal P}^{ef,{\bar e}{\bar f}} {\cal P}^{gh,{\bar g}{\bar h}} \, ,
\ea
where $K=P+Q$ and the pre-factor $1/2$ is the symmetry factor. In the above equation, all the color indices except $a$, $b$, $c$ and $d$ need to be summed. In addition, we use ${\cal V}_{{\rm L}/{\rm R}}$ to denote the Feynman rules for the left and right vertex, respectively. Using the following identities
\be
\sum_{{\bar e}{\bar f}} f^{ab,{\bar e}{\bar f},gh}  {\cal P}^{ef,{\bar e}{\bar f}}= f^{ab,fe,gh}\, ,\quad \sum_{{\bar g}{\bar h}} f^{cd,{\bar g}{\bar h},ef}  {\cal P}^{gh,{\bar g}{\bar h}}= f^{cd,hg,ef}\, ,
\ee
where the two structure constants $f^{ab,{\bar e}{\bar f},gh}$ and $f^{cd,{\bar g}{\bar h},ef}$ come from the left and right vertex, respectively, the sum over color indices ${\bar e}$, ${\bar f}$, ${\bar g}$ and ${\bar h}$ can be carried out. In terms of the retarded/advanced and symmetric bare propagators, Eq.~(\ref{pirglue}) can be expressed as
\ba\la{pirgluenew}
\Pi^{(1)ab,cd}_{R;\mu \nu}(P,\cQ)
&=& \frac{i}{4} g^2 \sum_{{\rm colors}}\int\frac{d^4 K}{(2\pi)^4} \big[D_{F}(Q,\cQ^{gh})D_{R}(K)+D_{A}(Q)D_{F}(K,\cQ^{ef})\big]\,\nonumber \\
&\times&f^{ab,fe,gh}f^{cd,hg,ef} {\cal L}^{(1)} _{\mu\nu}(P,Q,K)\, ,
\ea
where the Lorentz structure is fully contained in ${\cal L} ^{(1)}_{\mu\nu}(P,Q,K)$. Since it has nothing to do with the background field, a straightforward calculation gives that
\ba
{\cal L} ^{(1)}_{\mu\nu}(P,Q,K)&=&\big[(P+K)_\rho g_{\mu \sigma}+(-K-Q)_\mu g_{\rho \sigma}+(Q-P)_\sigma g_{\rho \mu} \big] g^{\sigma \sigma^\prime} \nonumber\\
&\times&\big[(-P+Q)_{\sigma^\prime} g_{\nu \rho^\prime}+(-Q-K)_\nu g_{\rho^\prime \sigma^\prime}+(K+P)_{\rho^\prime} g_{\sigma^\prime \nu} \big] g^{\rho \rho^\prime}\nonumber \\
&=&\big[(P+K)^2+(P-Q)^2\big]g_{\mu\nu}+10 K_\mu K_\nu-5(K_\mu P_\nu+K_\nu P_\mu)-2 P_\mu P_\nu
\, .\nonumber \\
\ea

Turning to the ghost-loop diagram [see Fig.~\ref{gse} (c)], the corresponding calculation is very similar to the above.  It is easy to show the following:
\ba\la{pirglue2}
\Pi^{(2)ab,cd}_{R;\mu \nu}(P,\cQ)
&=& - \frac{i}{4} g^2 \sum_{{\rm colors}}\int\frac{d^4 K}{(2\pi)^4} \big[D_{F}(Q,\cQ^{gh})D_{R}(K)+D_{A}(Q)D_{F}(K,\cQ^{ef})\big]\,\nonumber \\
&\times&f^{ab,fe,gh}f^{cd,hg,ef} 2K_\mu Q_\nu\, .
\ea
In order to get a final result as given by Eq.~(\ref{pirglu}), we need to rewrite the above equation with a more symmetric form. By changing the variables $Q\rightarrow -K^\prime$ and $K\rightarrow -Q^\prime$, we can show that
\ba\la{change}
&&\int\frac{d^4 K}{(2\pi)^4} \big[D_{F}(Q,\cQ^{gh})D_{R}(K)+D_{A}(Q)D_{F}(K,\cQ^{ef})\big] K_\mu Q_\nu \nonumber \\
&&= \int\frac{d^4 Q^\prime}{(2\pi)^4} \big[D_{F}(-K^\prime,\cQ^{gh})D_{R}(-Q^\prime)+D_{A}(-K^\prime)D_{F}(-Q^\prime,\cQ^{ef})\big] Q^\prime_\mu K^\prime_\nu \nonumber \\
&&= \int\frac{d^4 K^\prime}{(2\pi)^4} \big[D_{F}(K^\prime,\cQ^{hg})D_{A}(Q^\prime)+D_{R}(K^\prime)D_{F}(Q^\prime,\cQ^{fe})\big] Q^\prime_\mu K^\prime_\nu\, ,
\ea
where we changed $d^4 Q^\prime$ into $d^4 K^\prime$ because $K^\prime=P+Q^\prime$. In addition, under the sign change of the momentum in the symmetric propagator, the flip of the color indices in the last line in Eq.~(\ref{change}) is due to the definition of the distribution function as given by Eq.~(\ref{bd}). By realizing the fact that the product of the two structure constants in Eq.~(\ref{pirglue2}) is invariant under the interchanges of color indices, $h\leftrightarrow e$ and $g\leftrightarrow f$, it can be shown that
\ba\la{pirglue2new}
\Pi^{(2)ab,cd}_{R;\mu \nu}(P,\cQ)
&=& \frac{i}{4} g^2 \sum_{{\rm colors}}\int\frac{d^4 K}{(2\pi)^4} \big[D_{F}(Q,\cQ^{gh})D_{R}(K)+D_{A}(Q)D_{F}(K,\cQ^{ef})\big]\,\nonumber \\
&\times&f^{ab,fe,gh}f^{cd,hg,ef} {\cal L} ^{(2)}_{\mu\nu}(P,Q,K)\, ,
\ea
where
\be
{\cal L} ^{(2)}_{\mu\nu}(P,Q,K)=- (2K_\mu Q_\nu+2 Q_\mu K_\nu)/2=- 2K_\mu K_\nu+P_\mu K_\nu+K_\mu P_\nu\, .
\ee

Finally, we consider the tadpole diagram [see Fig.~\ref{gse} (d)] which can be expressed as
\be\la{pirglue3}
\Pi^{(3)ab,cd}_{R;\mu \nu}(P,\cQ)
=  \frac{3}{4} g^2 \sum_{{\rm colors}}\int\frac{d^4 K}{(2\pi)^4} D_{F}(K,\cQ^{ef}) (f^{ab,ef,ij}f^{cd,ji,fe}-f^{ab,fe,ij}f^{cd,ef,ji}) g_{\mu \nu}\, .
\ee
Notice that one of the three terms in the four-gluon vertex (see the Feynman rules in Fig.~\ref{fey}) vanishes because $\sum_{gh}f^{ij,ef,gh}{\cal P}^{ef,gh}=0$ where the projection operator comes from the gluon propagator. Clearly, the above equation needs to be rewritten in a form analogous to the corresponding results from the other two Feynman diagrams. To do so, we first express Eq.~(\ref{pirglue3}) in the following form:
\be\la{pirglue3new}
\Pi^{(3)ab,cd}_{R;\mu \nu}(P,\cQ)
=  \frac{3}{4} g^2 \sum_{{\rm colors}}\int\frac{d^4 K}{(2\pi)^4} \big[D_{F}(Q,\cQ^{fe})+D_{F}(K,\cQ^{ef})\big] f^{ab,fe,ij}f^{cd,ji,ef} g_{\mu \nu}\, .
\ee
To get the above equation, the color indices $e$ and $f$ in $D_{F}(K,\cQ^{ef}) f^{ab,ef,ij}f^{cd,ji,fe}$ in Eq.~(\ref{pirglue3}) have been interchanged, and the resulting $D_{F}(K,\cQ^{fe})$ is then replaced by $D_{F}(Q,\cQ^{fe})$ which is valid under the integral $\int d^4 K$. Furthermore, for the contribution associated with $D_{F}(Q,\cQ^{fe})$, we can insert a term $-i D_R(K) (K^2+Q^2)$ into Eq.~(\ref{pirglue3new}) because such an inserted term is effectively ``$1$ " under the integral thanks to the delta function in the propagator. Similarly, a term $-i D_A(Q) (K^2+Q^2)$ can be also inserted into the above equation for the contribution associated with $D_{F}(K,\cQ^{ef})$. Consequently, we arrive at
\ba\la{tadpole}
\Pi^{(3)ab,cd}_{R;\mu \nu}(P,\cQ)
&=&  \frac{i}{4} g^2 \sum_{{\rm colors}}\int\frac{d^4 K}{(2\pi)^4} \big[D_{F}(Q,\cQ^{gh})D_{R}(K)+D_{A}(Q)D_{F}(K,\cQ^{ef})\big]\,\nonumber \\
&\times&f^{ab,fe,gh}f^{cd,hg,ef} {\cal L} ^{(3)}_{\mu\nu}(P,Q,K)\, ,
\ea
with ${\cal L} ^{(3)}_{\mu\nu}(P,Q,K)=-3(K^2+Q^2)g_{\mu\nu}$.

Summing up the above results, the total contribution from the pure gauge part to the retarded gluon self-energy reads
\ba\la{pirgluetot}
\Pi^{ab,cd}_{R;\mu \nu}(P,\cQ)
&=& \frac{i}{4} g^2 \sum_{{\rm colors}}\int\frac{d^4 K}{(2\pi)^4} \big[D_{F}(Q,\cQ^{gh})D_{R}(K)+D_{A}(Q)D_{F}(K,\cQ^{ef})\big]\,\nonumber \\
&\times&f^{ab,fe,gh}f^{cd,hg,ef} {\cal L} _{\mu\nu}(P,Q,K)\, ,
\ea
where
\ba
 {\cal L}_{\mu\nu}(P,Q,K)&=& {\cal L} ^{(1)}_{\mu\nu}(P,Q,K)+ {\cal L} ^{(2)}_{\mu\nu}(P,Q,K)+ {\cal L} ^{(3)}_{\mu\nu}(P,Q,K)\nonumber \\
 &\approx& 8K_\mu K_\nu-4K_\mu P_\nu-4P_\mu K_\nu -4 Q\cdot K g_{\mu\nu}\, .
\ea
Here, we drop terms $\sim P_\mu P_\nu$ which give contributions beyond NLO in the HTL approximation. At this point, it is clear to see Eq.~(\ref{pirgluetot}) is identical to Eq.~(\ref{pirglu}).

Given the above calculations, the corresponding contributions to the symmetric gluon self-energy from Figs.~\ref{gse} (b), \ref{gse} (c) and \ref{gse} (d) can be obtained straightforwardly. Notice that the tadpole diagram does not contribute to $\Pi^{ab,cd}_{F;\mu \nu}(P,\cQ)$. However, in order to show its similarity with Eq.~(\ref{tadpole}), we can write such a zero result as
\ba
\Pi^{(3)ab,cd}_{F;\mu \nu}(P,\cQ)
&=&  \frac{i}{4} g^2 \sum_{{\rm colors}}\int\frac{d^4 K}{(2\pi)^4} \big[D_{F}(Q,\cQ^{gh})D_{F}(K,\cQ^{ef})-\big(D_{R}(K)-D_{A}(K)\big)\,\nonumber \\
&\times&\big(D_{R}(Q)-D_{A}(Q)\big)\big] f^{ab,fe,gh}f^{cd,hg,ef} (-3) (K^2+Q^2)g_{\mu\nu}\, .
\ea
The above equation vanishes due to a product of delta functions $\delta(K^2)\delta(Q^2)$ from the propagators. As a result, the symmetric gluon self-energy $\Pi^{(i)ab,cd}_{F;\mu \nu}(P,\cQ)$ with $i=1,2,3$ can be simply obtained from the retarded solutions $\Pi^{(i)ab,cd}_{R;\mu \nu}(P,\cQ)$ as given in Eqs.~(\ref{pirgluenew}), (\ref{pirglue2new}) and (\ref{tadpole}) by the following replacement:
\ba
&& D_{F}(Q,\cQ^{gh})D_{R}(K)+D_{A}(Q)D_{F}(K,\cQ^{ef})\nonumber \\
&&\quad \longrightarrow D_{F}(Q,\cQ^{gh})D_{F}(K,\cQ^{ef})-\big(D_{R}(K)-D_{A}(K)\big)\big(D_{R}(Q)-D_{A}(Q)\big)\, .
\ea
Thus, up to NLO in the HTL approximation, the three diagrams in the pure gauge part lead to the following contributions to the symmetric gluon self-energy
\ba\la{pifglu}
&&\Pi^{ab,cd}_{F;\mu \nu}(P,\cQ) = i g^2\sum_{efgh} f^{ab,fe,gh} f^{cd,hg,ef} \int\frac{d^4 K}{(2\pi)^4}  (2 K_\mu K_\nu-P_\mu K_\nu-K_\mu P_\nu-g_{\mu\nu} Q \cdot K)\,\nonumber \\
&&\quad \times\big[D_{F}(Q,\cQ^{gh})D_{F}(K,\cQ^{ef})-\big(D_{R}(K)-D_{A}(K)\big)\big(D_{R}(Q)-D_{A}(Q)\big)\big]\,.
\ea
As we can see, the above equation is analogous to the quark-loop contribution to $\Pi^{ab,cd}_{F;\mu \nu}(P,\cQ)$ which is given by Eq.~(\ref{pif}).

\section{ REORGANIZATION  OF  THE LO SYMMETRIC  GLUON  SELF-ENERGY}\la{repif}

As shown in Eqs.~(\ref{pir0}) and (\ref{pif0}), both the NLO $\Pi_{R/A}$ and the LO $\Pi_F$ can reproduce the correct limit at $\cQ=0$. For nonzero background field, however, they get different modifications on the Debye mass. In fact, such a background field modification originates from the integral over $k$ where the integrand of the retarded solution consists of a sum of two distribution functions, see Eq.~(\ref{pirnlo}). On the other hand, according to Eq.~(\ref{pif3}), the corresponding integral in the symmetric solution involves a more complicated combination of the distribution functions. In order to show more similarities between the two, it appears to be interesting to artificially introduce a term to the LO symmetric gluon self-energy which vanishes at $\cQ=0$, and thus gives the anomalous contribution. Subtracting it from Eq.~(\ref{piflofinal}), we can define the so-called normal contribution. 

Taking the fermionic part [Eq.~(\ref{pif3})] as an example, explicitly, we can write
\ba\la{pifano}
\Pi^{\rm ano}_{F;\mu \nu}(P,\cQ)&=&- i  \frac{g^2}{4\pi^2 p} N_f \delta^{ad}\delta^{bc} \int k^2 dk  [( {\tilde n}_+^a(k)- {\tilde n}_+^c(k))^2+( {\tilde n}_-^a(k)- {\tilde n}_-^c(k))^2] \Lambda_{\mu\nu}(P)\, \nonumber \\
&=& i \frac{T}{ p} m_f^2\delta^{ad}\delta^{bc}\big[{\cal F}_f^{(1)}({\sf q}^a,{\sf q}^c)+\frac{1}{2}\cot(\pi {\sf q}^{ac}){\cal G}_f({\sf q}^a,{\sf q}^c)\big]\Lambda_{\mu\nu}(P)\, ,\
\ea
and
\ba\la{pifnor}
&&\Pi^{\rm nor}_{F;\mu \nu}(P,\cQ)= i  \frac{g^2 T}{4\pi^2 p} N_f \int k d k \Big\{\delta^{ab}\delta^{cd}\frac{4}{N} \Big[{\tilde n}_+^a(k)+{\tilde n}_-^a(k) +{\tilde n}_+^c(k)+{\tilde n}_-^c(k)\,\nonumber\\
&&\quad-\frac{1}{N}\sum_e({\tilde n}_+^e(k)+{\tilde n}_-^e(k))\Big]
- 2 \delta^{ad}\delta^{bc}  \big({\tilde n}_+^a(k)+{\tilde n}_-^a(k) +{\tilde n}_+^c(k)+{\tilde n}_-^c(k)\big)\Big\}  \Lambda_{\mu\nu}(P)\, \nonumber \\
&&\quad = i\frac{T}{p} m_f^2 \Big[\delta^{ab} \delta^{cd} \frac{1}{N}{\cal F}^{(2)}_f({\sf q}^a,{\sf q}^c)-\delta^{ad} \delta^{bc}  {\cal F}^{(1)}_f({\sf q}^a,{\sf q}^c)\Big]  \Lambda_{\mu\nu}(P)\, .
\ea

The above anomalous contribution shares some similarities with $\Pi^{\rm ano}_{R;\mu \nu}(P,\cQ)$ as given in Eq.~(\ref{pirtotlo}). They both come from the LO terms in the HTL approximation and contain only the off-diagonal components. In addition, the anomalous contribution vanishes when $\cQ=0$ or ${\sf q}^a={\sf q}^c$ because a difference between two distribution functions appears in the integrand in Eq.~(\ref{pifano}). Similar as before,  to avoid an  ``$\infty \cdot0$ " ambiguity in our analytical result, ${\sf q}^a={\sf q}^c$ should be understood as ${\sf q}^a\rightarrow {\sf q}^c \pm \epsilon$. Then the second line in Eq.~(\ref{pifano}) equals zero as expected. However, different from Eq.~(\ref{pirtotlo}), such an artificially introduced anomalous contribution is orthogonal to $P^\mu$ and does not satisfy a similar relation as given by Eq.~(\ref{idsum}).

Although it is trivial to show that the sum of Eqs.~(\ref{pifano}) and (\ref{pifnor}) is identical to Eq.~(\ref{piflofer}), the motivation for separating Eq.~(\ref{pif3}) into the above two parts becomes clear when we look at the normal contributions. 
According to Eqs.~(\ref{pirtotnlo}) and (\ref{pifnor}), the background field modification on $m_f^2$ which involves only $B_2(x)$ is exactly the same for both the retarded/advanced and symmetric gluon self-energies. Consequently, the $\cQ$-dependent mass squared as defined in Eq.~(\ref{qmd}) is universal to describe the background field modification on the normal contributions in all the three independent real-time gluon self-energies. Furthermore, it also turns out to be meaningful when we include nonperturbative contributions in the gluon self-energy which however, is beyond the scope of current work and we only provide a very preliminary discussion at the end of Sec.~\ref{conclusions}.

A similar consideration also applies to the pure gauge part and the corresponding anomalous and normal contributions are given by the following equations:
\ba\la{pifano2}
\Pi^{\rm ano}_{F;\mu \nu}(P,\cQ)
&=& i \frac{T}{ p} m_g^2\delta^{ad}\delta^{bc}\big[{\cal F}_g^{(1)}({\sf q}^a,{\sf q}^c)-\frac{1}{2}\cot(\pi {\sf q}^{ac}){\cal G}_g({\sf q}^a,{\sf q}^c)\big]\Lambda_{\mu\nu}(P)\, ,\\ \la{pifnor2}
\Pi^{\rm nor}_{F;\mu \nu}(P,\cQ)
&=& i\frac{T}{p} m_g^2 \Big[\delta^{ab} \delta^{cd}\frac{1}{N} {\cal F}^{(2)}_g({\sf q}^a,{\sf q}^c)-\delta^{ad} \delta^{bc}  {\cal F}^{(1)}_g({\sf q}^a,{\sf q}^c)\Big]  \Lambda_{\mu\nu}(P)\, .
\ea

Finally, we can also express the normal contributions of the LO symmetric gluon self-energy as
\be\la{effpif}
\Pi^{\rm nor}_{F;\mu \nu}(P,\cQ)=- i\frac{T}{p}({\cal M}_D^2)^{ab,cd}(\cQ) \Lambda_{\mu\nu}(P) \, ,
\ee
where $({\cal M}_D^2)^{ab,cd}(\cQ)$ has been already defined in Eq.~(\ref{qmd}).



\end{document}